\documentclass[11pt,letterpaper]{article}
\usepackage[margin=1in]{geometry}
\usepackage{amsmath,amssymb}
\usepackage[authoryear,round]{natbib}
\usepackage{graphicx,longtable,needspace,placeins,float}
\usepackage{url,xcolor}
\usepackage[hidelinks]{hyperref}
\numberwithin{equation}{section}
\newcommand{\cmatgraphic}[2]{#2}
\newcommand{\paperfigure}[2]{\FloatBarrier\begin{figure}[H]\centering\includegraphics[width=\linewidth]{figures/figure_#1.pdf}\caption{#2}\label{fig:method-#1}\end{figure}}
\newcommand{\expfigure}[2]{\FloatBarrier\begin{figure}[H]\centering\cmatgenerationgraphic{#1}\caption{#2}\label{fig:exp-#1}\end{figure}}
\newcommand{\cmatgenerationgraphic}[1]{%
\begingroup
\setlength{\unitlength}{1bp}%
\normalfont\fontfamily{ptm}\fontsize{8.5bp}{10bp}\selectfont
\ifnum#1=8\relax
  \resizebox{\linewidth}{!}{\begin{picture}(404.5,210)
    \put(0,0){\includegraphics[width=404.5bp]{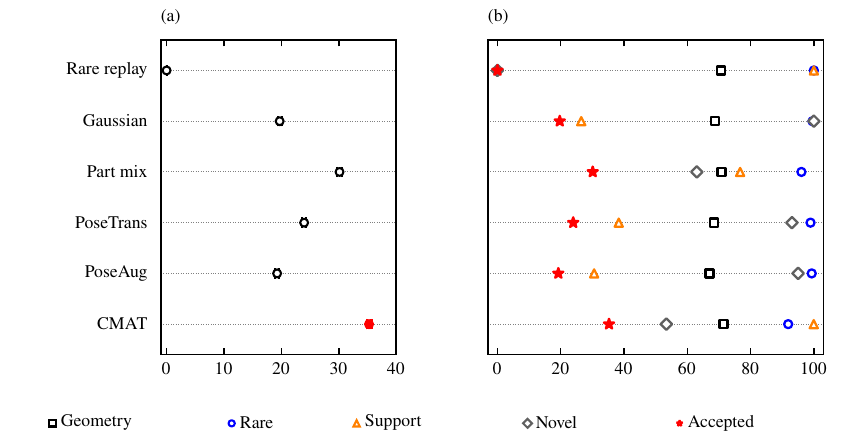}}
    \put(3.8,96.9){\color{white}\rule{68bp}{10.6bp}}
    \put(70.809,100.551){\makebox[0pt][r]{PoseTrans-inspired}}
    \put(20,73.4){\color{white}\rule{51.8bp}{10.6bp}}
    \put(70.8,77.029){\makebox[0pt][r]{PoseAug-style}}
  \end{picture}}%
\else\ifnum#1=9\relax
  \resizebox{\linewidth}{!}{\begin{picture}(404.5,277)
    \put(0,0){\includegraphics[width=404.5bp]{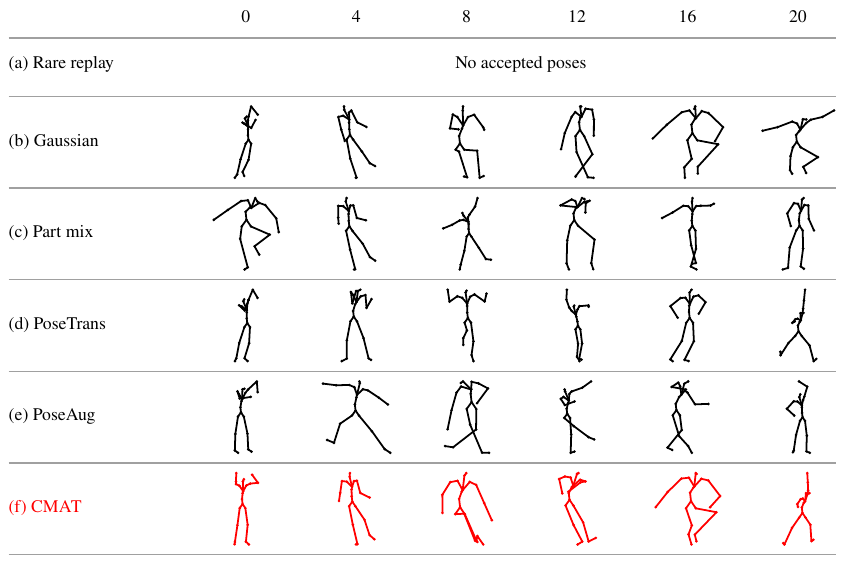}}
    \put(3.2,115.1){\color{white}\rule{80bp}{10.6bp}}
    \put(4.199,118.849){\makebox[0pt][l]{(d) PoseTrans-inspired}}
    \put(3.2,71.3){\color{white}\rule{64bp}{10.6bp}}
    \put(4.199,75.053){\makebox[0pt][l]{(e) PoseAug-style}}
  \end{picture}}%
\else
  \includegraphics[width=\linewidth]{figures/figure_#1.pdf}%
\fi\fi
\endgroup}
\newcounter{algorithm}
\newcounter{algline}
\newenvironment{algorithm}[1]{%
\refstepcounter{algorithm}\setcounter{algline}{0}\def\algtitle{#1}%
\begingroup\small\setlength{\parskip}{0pt}%

\let\refstepcounter\stepcounter
\begin{longtable}{@{}r@{\quad}p{0.91\linewidth}@{}}
\hline\multicolumn{2}{@{}l}{\textbf{Algorithm \thealgorithm\quad #1}}\\\hline\endfirsthead
\hline\multicolumn{2}{@{}l}{\textbf{Algorithm \thealgorithm\quad #1 (continued)}}\\\hline\endhead
\hline\endfoot
}{\end{longtable}\addtocounter{table}{-1}\endgroup}
\newcommand{\AL}[2][0]{\stepcounter{algline}\thealgline & \hspace*{#1em}\begin{minipage}[t]{\dimexpr\linewidth-#1em\relax}\raggedright #2\strut\end{minipage}\\}
\newcommand{\AI}[2]{ & \textbf{#1}: #2\\}

\begin{document}
\title{Local Manifold Attention for Recovery and Generation of Rare 3D Human Poses}
\author{%
Feiyue Pan\textsuperscript{1}\quad
Zhigang Yao\textsuperscript{2,*}\\[0.8em]
\begin{minipage}{0.96\textwidth}\small\centering
\textsuperscript{1}School of Mathematical Sciences, Fudan University;\\
Shanghai Institute for Mathematics and Interdisciplinary Sciences,\\
Shanghai, China.\\[0.4em]
\textsuperscript{2}Department of Statistics and Data Science,\\
National University of Singapore, Singapore\\[0.4em]
\textsuperscript{*}Corresponding author:
\href{mailto:zhigang.yao@nus.edu.sg}{zhigang.yao@nus.edu.sg}
\end{minipage}}
\date{}
\maketitle

\begin{abstract}
A rare 3D human pose in a motion sequence may reflect either a genuine movement or an estimation error, making correction difficult when reference databases contain few comparable examples. We investigate whether local geometric structure can support the recovery and generation of rare configurations, using motion-capture sequences from AMASS, which unifies datasets through a common parametric human-body model, together with video-based pose estimates. We develop the cylindrical manifold attention transformer (CMAT), which extends local manifold fitting through geometry- and task-dependent neighborhood weights. For motion correction without a pretrained correction network, Self-Action CMAT retrieves comparable windows from within a sequence, adjusts their contributions for neighborhood imbalance and temporal dependence, and integrates local geometry with robust temporal regularization. Experiments on perturbed and estimated motions show reductions in position and acceleration errors, with component analyses examining how weighting and temporal processing contribute to recovery. The controlled correction comparisons are exploratory and conditional on the selected motion-amplitude bound. For generation, Rarity-preserving CMAT explores locally supported configurations subject to independently calibrated rarity and separation from stored observations. Under common screening criteria, it achieves a higher yield of eligible rare poses than the evaluated generation baselines.
\end{abstract}

\raggedbottom

\FloatBarrier
\section{Introduction}\label{sec:introduction}

\FloatBarrier
\subsection{Background and related work}\label{sec:background}

{Human motion sequences describe changes in joint positions and orientations over time, but estimated trajectories may contain spatial errors and temporal jitter. {This makes correction of rare motions particularly challenging: denoising must both reduce these artifacts and preserve genuine motion variation.}}

{We define rarity relative to a specified reference motion database: a whole-body configuration is rare when it has few similar observations, but low frequency alone does not imply an estimation error. For example, arm raising and torso rotation may each be common, while their combination with a particular leg configuration occurs less often. Correcting such a pose solely because of its low population frequency risks distorting valid motion, especially when nearby phases, repeated movements or coordinated joint changes within the same sequence provide local evidence of its structure.}

{To distinguish global rarity from local motion structure, we use standardized motion-capture representations.} {The SMPL parametric body model \citep{loper2015smpl} maps joint rotations and body-shape parameters to the body surface and joint positions, and the AMASS archive \citep{mahmood2019amass} unifies multiple motion-capture datasets in this representation,} {enabling direct comparisons of whole-body configurations across sequences while preserving the temporal order of each recording.}

Existing methods learn pose or motion priors across sequences to supplement unreliable observations. {At the single-pose level, Pose-NDF \citep{tiwari2022posendf} learns pose geometry for assessing and correcting departures from supported configurations, whereas NRDF \citep{he2024nrdf} models distances and projections in joint-rotation space. For continuous motion, HuMoR \citep{rempe2021humor} incorporates learned motion regularities into sequence optimization, while RoHM \citep{zhang2024rohm} recovers trajectories affected by noise or occlusion using a learned motion model.} {The usefulness of these priors for {rare} configurations depends on their coverage in the training data: because consecutive frames are correlated, even a large archive may represent only a narrow range of poses and provide little information about rare movements.}

{For trajectory denoising, we can also draw on relationships between frames in the observed sequence. In video-based estimation, VIBE \citep{kocabas2020vibe} combines temporal modeling with a motion prior, whereas SmoothNet \citep{zeng2022smoothnet} uses longer-range temporal relationships to reduce jitter in trajectories that have already been estimated. When applying temporal correction to rare motions, however, we must distinguish estimation noise from genuine rapid changes to avoid smoothing away the movement being recovered.}

{We also study the synthesis of new three-dimensional poses in sparsely represented regions of a reference database. Starting from rare reference poses and their neighborhoods, we seek configurations that remain rare, differ from stored observations and have local geometric support. Related augmentation strategies include PoseAug \citep{gong2021poseaug}, which learns pose transformations from a three-dimensional estimator's error feedback, and PoseTrans \citep{jiang2022posetrans}, which transforms images and two-dimensional keypoints to increase coverage of low-frequency configurations. For direct three-dimensional generation, however, local support, rarity relative to the reference distribution and separation from stored samples must hold together: replaying observations cannot expand geometric coverage, whereas unconstrained perturbations may lack structural support. We therefore evaluate generators by the fraction of all attempts that jointly satisfy these requirements.}

{Despite their different objectives, rare-motion correction and rare-pose generation face a common finite-sample problem: neighboring observations may be unevenly distributed and dependent, making local geometric estimates unreliable.} When population coverage is limited, can geometric relationships among local observations support both the recovery of rare motions and the exploration of supported new configurations? Manifold fitting offers a way to organize this local evidence by estimating a smooth, lower-dimensional structure in the original observation space from noisy samples. \citet{yao2025manifold} study recovery under unbounded Gaussian noise using tangent-space estimates at latent projected locations. \citet{yao2023manifold} develop a two-step construction that first estimates a local contraction direction and then takes a weighted local average within a cylinder aligned with that direction. These developments provide a statistical basis for recovering geometry from noisy observations under explicit sampling and noise assumptions.

Manifold fitting has also been used in settings where analysis depends on recovering geometric relationships among high-dimensional observations. For example, \citet{yao2024cyclegan} learn smooth mappings between latent and ambient spaces to support manifold estimation, projection and nonlinear interpolation. In single-cell analysis, \citet{yao2024singlecell} use manifold fitting to denoise RNA-sequencing data and refine cell clustering and visualization. \citet{li2025metabolomic} use fitted manifolds to characterize metabolomic heterogeneity and examine disease associations in UK Biobank populations. These applications illustrate how estimating geometry can support denoising, interpolation and the analysis of heterogeneous populations.

Comparable motion phases and neighboring poses may provide local evidence of joint coordination even when a configuration is globally rare. We develop the cylindrical manifold attention transformer (CMAT) to use this evidence for correcting rare trajectories and generating new, locally supported poses.

\FloatBarrier
\subsection{Main contributions}\label{sec:contributions}

{Our main contributions are as follows:}

\begin{itemize}
\item We extend neighborhood weighting in cylindrical local fitting to construct the cylindrical manifold attention transformer (CMAT). While retaining the geometric support of the cylindrical neighborhood, we allow neighbors' contributions to vary with local geometry, observation reliability and relationships among observations, and select or combine local estimates according to the task.
\item We develop two task-specific implementations with different sources of information. Self-Action CMAT retrieves comparable windows from the current sequence, including similar nonadjacent motion phases. These windows may share errors and be unevenly distributed along the local structure, motivating neighborhood weighting and local-constraint selection to adjust their contributions. Combined with robust temporal regularization, this yields joint-trajectory correction without a pretrained correction network. Rarity-preserving CMAT instead draws local support from neighboring poses in the reference database. It explores new configurations while constraining independently calibrated rarity and separation from stored poses, to avoid concentrating on common configurations or merely reproducing existing observations.
\end{itemize}

{In the controlled AMASS evaluation, Self-Action CMAT achieves lower errors than the evaluated comparators within the selected evaluation population.}

For generation, under common screening criteria, Rarity-preserving CMAT increases the yield of samples meeting requirements for local support, rarity and separation from stored poses relative to the evaluated generation baselines.

\begin{figure}[H]
\centering
\includegraphics[width=\textwidth]{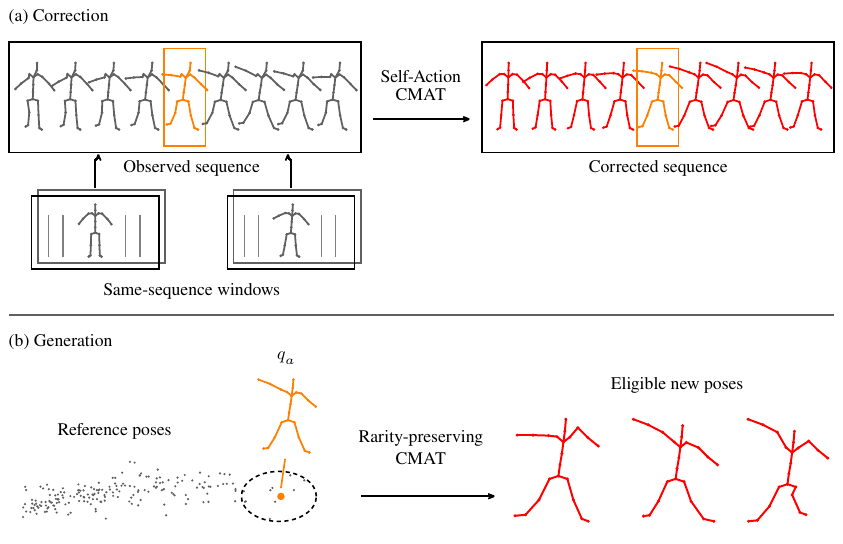}
\caption{{Two uses of local geometry for rare-motion recovery and rare-pose generation.}}
\label{fig:study-overview}
\end{figure}

\Needspace{30\baselineskip}
{Figure~\ref{fig:study-overview} summarizes each task's input, local information source and output.}

\FloatBarrier

\FloatBarrier
\subsection{Outline}\label{sec:outline}

{We first develop the representations, rarity calibration and CMAT algorithms in Section~2, then present the experiments in Section~3 before discussing their findings and limitations in Section~4. Full settings, computational details and supplementary analyses appear in Sections A--F of the Supplementary Material \citep{cmat2026supp}.}

\Needspace{8\baselineskip}
\FloatBarrier
\Needspace{8\baselineskip}
\section{Methods}\label{sec:methods}

\FloatBarrier
\subsection{Data and pose representation}\label{sec:data}
{AMASS unifies records from multiple optical motion-capture datasets into parametric motion sequences through body-model fitting. These sequences contain per-frame poses, global translations, body shapes and sampling rates, and can be used to reconstruct time-varying three-dimensional human surfaces (Figure~\ref{fig:method-2}(a)).}

{For a single human pose, we use the local rotations of 21 body joints to describe each joint's orientation relative to its parent. As shown in Figure~\ref{fig:method-2}(b), the rotation of joint $j$ at frame $t$ is represented by a three-dimensional axis-angle vector. A pose can therefore be stored as an array of axis-angle vectors or flattened to 63 coordinates:}
\begin{equation}\begin{gathered}
q_t=(\boldsymbol r_{t,j}^\top)_{j=1}^{21}\in\mathbb R^{21\times3},\qquad
\boldsymbol r_{t,j}\in\mathbb R^3,\\
X=(q_1,\ldots,q_L),\qquad
\mathcal R=\{q^{(1)},\ldots,q^{(N)}\}.
\end{gathered}\label{eq:data-objects}\end{equation}
Here $X$ retains temporal order, whereas $\mathcal R$ contains sampled individual poses for reference. The direction and length of a nonzero axis-angle vector give the rotation axis and angle; its components are not Euler angles.

We identify a pose $q$ with its rotations $R_j=\operatorname{Exp}(\boldsymbol r_j)\in\mathrm{SO}(3)$, so that $q$ lies in $\mathrm{SO}(3)^{21}$. For two poses, the principal relative rotation vectors and their concatenation are
\begin{equation}
\boldsymbol\xi_j=\operatorname{Log}(R_j^\top R'_j),\qquad
\operatorname{Log}_q(q')=(\boldsymbol\xi_1^\top,\ldots,\boldsymbol\xi_{21}^\top)^\top\in\mathbb R^{63}.
\label{eq:pose-log}
\end{equation}
Each $\|\boldsymbol\xi_j\|$ is the shortest angle between the corresponding joint orientations. We use the product rotation distance throughout:
\begin{equation}
d(q,q')=\|\operatorname{Log}_q(q')\|_2
=\left\{\sum_{j=1}^{21}\|\operatorname{Log}(R_j^\top R'_j)\|^2\right\}^{1/2}.
\label{eq:distance-scale}
\end{equation}
A local increment $\boldsymbol\delta=(\boldsymbol\delta_1^\top,\ldots,\boldsymbol\delta_{21}^\top)^\top$ is applied through
\begin{equation}
q^{\rm new}=\operatorname{Exp}_q(\boldsymbol\delta),\qquad
R_j^{\rm new}=R_j\operatorname{Exp}(\boldsymbol\delta_j),
\label{eq:pose-update}
\end{equation}
{after which we convert the updated rotations to axis-angle coordinates for storage and decode them through SMPL when joint positions or surface geometry are needed.}

\Needspace{40\baselineskip}
{Figure~\ref{fig:method-2} illustrates the correspondence between these pose representations and full-body configurations. In the two-dimensional projection in panel~(d), rare configurations appear in regions associated with different forms of motion variation, rather than only along the outer boundary of the embedding.}

\paperfigure{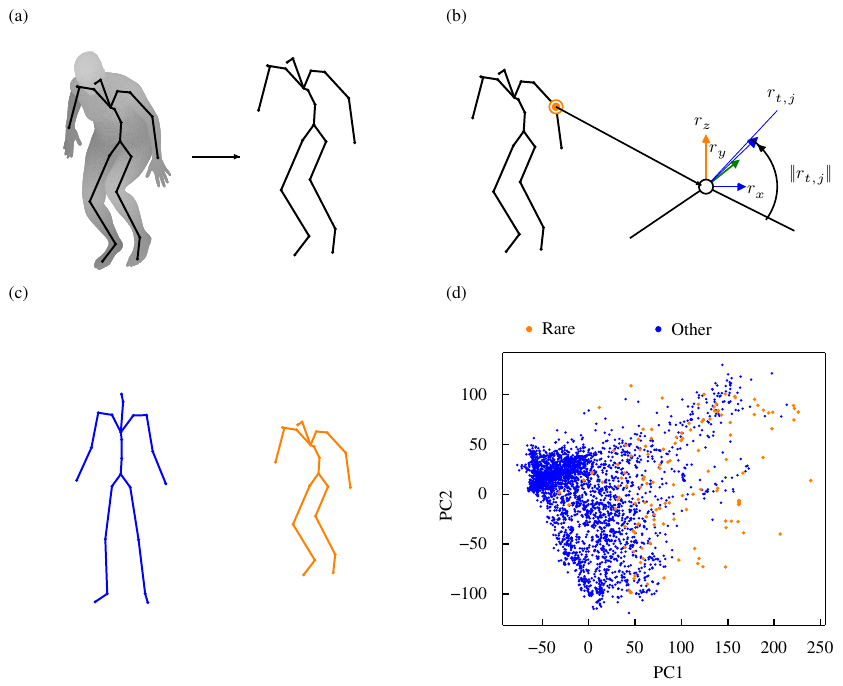}{{AMASS pose representation and rare configurations. (a) The SMPL surface and analysis joints of the same pose. (b) The 21 local body-joint rotations, represented by three-dimensional axis-angle vectors. (c) Ordinary (left) and rare (right) poses from the same motion. (d) {Poses randomly sampled from AMASS} projected from their 63-dimensional axis-angle features.}}

\FloatBarrier
\subsection{Calibrating rarity on independent subjects}\label{sec:rarity}
Rarity should describe limited proximity to different reference motions. For the retained poses $\mathcal R_s$ of reference sequence $s$, define
\begin{equation}
\delta_s(q)=\min_{q'\in\mathcal R_s}d(q,q'),\qquad
U_1(q)\leq U_2(q)\leq\cdots,
\label{eq:sequence-distances}
\end{equation}
where the $U_\ell$ are the ordered distances to distinct sequences. With $\mathcal K_{\rm seq}=\{4,8,16\}$, the multiscale score is
\begin{equation}
A_k(q)=\frac1k\sum_{\ell=1}^kU_\ell(q),\qquad
S_{\rm seq}(q)=\frac13\sum_{k\in\mathcal K_{\rm seq}}\log A_k(q).
\label{eq:sequence-score}
\end{equation}
{Because each $U_\ell(q)$ is based on the closest pose from a distinct sequence, a larger $S_{\rm seq}(q)$ indicates that the query remains distant from several reference recordings. Counting support by sequence prevents densely sampled adjacent frames from a single recording from artificially inflating that support.}

{We calibrate rarity using poses from subjects outside the reference library. Within collection $g$, let the calibration cohort contain $H_g$ subjects, with $L_h$ sampled sequences for subject $h$ and $m_{hs}$ selected poses in sequence $s$. We assign each pose weight $w_i=(H_gL_hm_{hs})^{-1}$ to give equal weight to subjects and, within each subject, to sequences. The resulting weighted calibration distribution and rarity percentile are}
\begin{equation}
\widehat G_g(t)=\sum_{i\in\mathcal C_g}w_i\mathbf1\{S_{\rm seq}(c_i)\leq t\},\qquad
\rho_g(q)=\widehat G_g\{S_{\rm seq}(q)\}.
\label{eq:independent-calibration}
\end{equation}
The operational definition of a rare pose is
\begin{equation}
\rho_g(q)\geq0.95.
\label{eq:independent-rare}
\end{equation}

\FloatBarrier
\subsection{Adding attention to local manifold fitting}\label{sec:local}\label{sec:manifold}
CMAT adds attention to local manifold fitting by assigning task-dependent weights to neighboring observations. In the present implementations, attention weights are determined by explicit rules or constrained optimization, without training a neural network. Following the terminology in the review by \citet{yao2026review}, we refer to the cylindrical local fitting operator introduced above as the Yao--Yau estimator. The original construction first estimates a direction from a local mean, then averages observations within a cylinder along that direction. CMAT retains this local support and changes how observations contribute to the second average.

\FloatBarrier
\subsubsection{The Yao--Yau estimator}
For query $x\in\mathbb R^p$ and observations $x_i$, the first step computes a weighted mean in the ball $B(x,r_0)$:
\begin{equation}\begin{aligned}
\widetilde\alpha_i(x)&=\left[1-\|x_i-x\|^2/r_0^2\right]_+^k,\quad
\alpha_i(x)=\frac{\widetilde\alpha_i(x)}{\sum_j\widetilde\alpha_j(x)},\\
F(x)&=\sum_i\alpha_i(x)x_i,\qquad k>2.
\end{aligned}\label{eq:yao-first}\end{equation}
Here $[t]_+=\max(t,0)$. The displacement $F(x)-x$ estimates a contraction direction. For nonempty support and $F(x)\ne x$, define
\begin{equation}
U_x=\frac{(F(x)-x)(F(x)-x)^\top}{\|F(x)-x\|^2},\quad
u_i=U_x(x_i-x),\quad v_i=(I-U_x)(x_i-x).
\label{eq:yao-axis}
\end{equation}
The second step uses the cylinder $V_x=\{x_i:\|u_i\|<H,\ \|v_i\|<R\}$ centered at $x$:
\begin{equation}
\beta_i(x)=\frac{w_{\parallel}(u_i)w_{\perp}(v_i)}{\sum_jw_{\parallel}(u_j)w_{\perp}(v_j)},\qquad
G_{\rm YY}(x)=\sum_i\beta_i(x)x_i.
\label{eq:yao-map}
\end{equation}
\Needspace{32\baselineskip}
The compactly supported kernels control axial and transverse contributions. The axial kernel stays constant for $\|u\|\leq H/2$ and then decays to zero at $H$; the transverse kernel decreases to zero at radius $R$. Their full formulas are retained in Supplementary Section~F. Figure~\ref{fig:fitting-intuition} illustrates the two local means and the resulting displacement.

\FloatBarrier\begin{figure}[H]
\centering
\includegraphics[width=\linewidth]{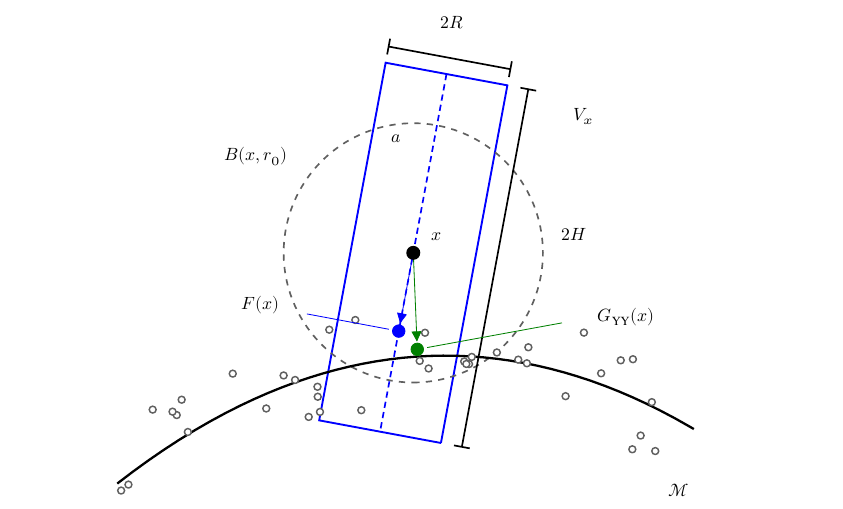}
\caption{{Two-dimensional illustration of the Yao--Yau estimator. Gray open circles represent observations near the manifold. The first weighted mean within a ball determines the cylinder axis; the second weighted mean within the cylinder gives the fitted output.}}
\label{fig:fitting-intuition}
\end{figure}

\FloatBarrier
\subsubsection{CMAT's extension and its task-specific use}
{CMAT retains the finite cylinder in the second stage of the Yao--Yau construction but replaces its fixed kernel-product weights with multiple attention heads, each applying a distinct weighting rule to the same local observations. For relative coordinates $\xi_i=x_i-x$, the common cylindrical support is}
\begin{equation}
\mathcal C(x)=\{i:x_i\in V_x\}.
\label{eq:cylinder-support-main}
\end{equation}
Head $h$ assigns score $L_{ih}(x)\in\mathbb R\cup\{-\infty\}$ and computes
\begin{equation}\begin{aligned}
\alpha_{ih}(x)&=\frac{\mathbf1\{i\in\mathcal C(x)\}\exp L_{ih}(x)}{\sum_{j\in\mathcal C(x)}\exp L_{jh}(x)},\\
h_h(x)&=\sum_i\alpha_{ih}(x)\xi_i,\qquad G_h(x)=x+h_h(x).
\end{aligned}\label{eq:token-attention}\end{equation}
{For each head, at least one neighbor within the cylinder receives a finite score, ensuring that the normalized weights are well defined; weights outside the cylinder are zero. We set the radius $R$ and half-length $H$ using the task-specific scales and neighborhood-expansion rules detailed in Supplementary Section~C.2.}

The original weighting is recovered with one head, the same cylinder and parameters, and
\begin{equation}
L_{i1}(x)=\log\{w_{\parallel}(u_i)w_{\perp}(v_i)\}
\label{eq:yao-attention-special}
\end{equation}
within the support. Then $\alpha_{i1}(x)=\beta_i(x)$ and $G_1(x)=G_{\rm YY}(x)$.

Heads may use explicit geometric scores or weights obtained by constrained optimization. Any nonnegative weights summing to one on the support can enter~\eqref{eq:token-attention} through their logarithms, with zero weights represented by $-\infty$.

\FloatBarrier
\subsection{Self-Action CMAT: correcting a sequence using its own motion information}\label{sec:correction}
{Let $Y\in\mathbb R^{L\times J\times3}$ contain joint positions after subtracting the root in each frame and dividing by one body scale for the sequence. We first fit a robust temporal guide to stabilize window retrieval and local geometric estimation, then combine the selected geometric constraints with the original observations in a joint optimization of the full trajectory.}

\FloatBarrier
\subsubsection{Roles of the temporal guide and the observations}
We first fit a robust temporal guide
\begin{equation}
G=\operatorname*{arg\,min}_Z\{\mathcal L_{\rm obs}(Z,Y)+\lambda_t\mathcal T(Z)\}.
\label{eq:correction-guide}
\end{equation}
{The robust observation loss is $\mathcal L_{\rm obs}(Z,Y)=\sum_{t,j}\psi_j(\|Z_{t,j}-Y_{t,j}\|)$, where $\psi_j(r)=\int_0^r u\omega_j(u)\,du$ and $\omega_j(r)=\max\{0.02,\min(1,2.8s_j/\max(r,10^{-12}))\}$. This loss downweights large joint-position residuals while retaining a minimum residual weight of 0.02. The temporal term is}
\begingroup
\begin{equation}
\mathcal T(Z)=\sum_{t=1}^{L-3}\sum_{j=1}^{J}
s_j\sqrt{\|(D^3Z)_{t,j}\|^2+(0.05s_j)^2},
\label{eq:correction-temporal}
\end{equation}
\endgroup
{where $D^3$ is the third temporal difference and $s_j$ is estimated from observed joint variation. This term regularizes abrupt changes in acceleration; the scale estimate and the guide implementation are specified in Supplementary Section~C.1.1.}

{We retain these two terms in the final objective~\eqref{eq:correction-combine}. We use the less variable guide coordinates $G$ for window retrieval and local geometric fitting, while retaining the original normalized observations $Y$ for observation--guide reliability comparisons, candidate scoring and the final data-fitting term.}

{Within overlapping time blocks, we retrieve comparable nine-frame windows using the guide's positions and first and second temporal differences, including both nearby times and nonadjacent recurrences of similar motion phases. Let $E_g$ extract and vectorize query window $g$, with $G_g=E_gG$. We express each matched window $E_iG=G_g+x_i$ in local coordinates $x_i=E_iG-G_g\in\mathbb R^{27J}$, then perform a rank-three preliminary fit and construct a finite cylindrical comparison set $\mathcal C_g$. We expand the cylinder as needed to retain at least $\min(24,K)$ of the $K\leq32$ retrieved matches, following the extraction and retrieval procedures in Supplementary Section~C.1.}

\FloatBarrier
\subsubsection{Neighborhood weights and geometric candidates}
{We use four weighting rules to account for distance differences, neighborhood imbalance and temporal dependence among matched windows. The first two use narrower and broader tangent-distance scales and downweight matches whose observations disagree with the guide. The third starts from the broader distance weights and penalizes a center lying to one side of the query in tangent coordinates, addressing uneven neighborhood placement. The fourth computes nonnegative generalized least-squares (GLS) weights under a working temporal correlation model, so that nearby matches sharing errors do not contribute as if they were equally independent. We denote the resulting weight vectors by $p^{(2)},p^{(8)},w^{\rm bal},w^{\rm GLS}$; their mathematical definitions and, where applicable, optimization procedures are given in Supplementary Section~C.1.3.}

For rule $a$, let $w_{gi}^{(a)}$ be nonnegative weights summing to one on the support. The local center is computed from the matched guide windows:
\begin{equation}
\mu_g^{(a)}=G_g+\sum_{i\in\mathcal C_g}w_{gi}^{(a)}x_i
=\sum_{i\in\mathcal C_g}w_{gi}^{(a)}E_iG.
\label{eq:correction-center-main}
\end{equation}
Principal directions are refitted under each weight vector. Four constraints then either retain variation along the first one, two or three principal directions, shrinking only their orthogonal complement toward the center, or shrink toward the full center with no direction left free. If $T_{ga,d}$ contains the first $d$ orthonormal principal directions, the projections are
\begin{equation}
\Pi_{g,(a,d)}=I-T_{ga,d}T_{ga,d}^{\top},\quad d=1,2,3,
\qquad \Pi_{g,(a,0)}=I.
\label{eq:correction-projections-main}
\end{equation}
{Crossing the four weighting rules with the four retained tangent dimensions $d\in\{0,1,2,3\}$ yields 16 local constraint candidates, indexed by $h=(a,d)$ with center $\mu_{gh}=\mu_g^{(a)}$ and projection $\Pi_{gh}$ onto the constrained directions. We attenuate corrections that are large relative to estimated residual variation through the coefficient $c_{gh}$, whose definition and associated variance estimate are provided in Supplementary Sections~C.1.3 and~C.1.4.}

\FloatBarrier
\subsubsection{Selecting local constraints relative to the guide}
{We compare each local geometric proposal with retaining the guide using a perturbation-based estimate of their risk difference. The proposal and its displacement from the guide are}
\begin{equation}\begin{aligned}
q_{gh}&=G_g+\frac{\eta_c c_{gh}}{1+\eta_c c_{gh}}\Pi_{gh}(\mu_{gh}-G_g),\\
\delta_{gh}(Y)&=q_{gh}(Y)-G_g(Y),\qquad Y_g=E_gY.
\end{aligned}\label{eq:candidate-score-input}\end{equation}
{Here $\eta_c$ controls geometric strength. With $D=27J$, we draw $N_\zeta\geq2$ independent full-sequence perturbations $\zeta_b\sim N(0,\widehat\Sigma)$ from the working covariance and denote their query-window components by $\zeta_{b,g}=E_g\zeta_b$. For perturbation size $\varepsilon>0$, we compute}
\begin{equation}\begin{split}
r_{bgh}={}&\frac{\|q_{gh}-Y_g\|^2-\|G_g-Y_g\|^2}{D}\\
&+\frac{\zeta_{b,g}^{\top}\{\delta_{gh}(Y+\varepsilon\zeta_b)-\delta_{gh}(Y-\varepsilon\zeta_b)\}}{\varepsilon D}.
\end{split}\label{eq:correction-risk-main}\end{equation}
{In~\eqref{eq:correction-risk-main}, the first term measures the per-coordinate change in squared residual against the observations, with negative values indicating closer agreement than the guide. The second term estimates an adjustment for the proposal's sensitivity to errors under the working dependence structure. To evaluate it, we recompute the guide, window matches and candidates on both perturbed sequences. }

For the query set $\mathcal G_B$ in time block $B$, define
\begin{equation}\begin{aligned}
R_{bBh}&=\frac1{|\mathcal G_B|}\sum_{g\in\mathcal G_B}r_{bgh},\\
\mathcal S_{Bh}&=\frac1{N_\zeta}\sum_{b=1}^{N_\zeta}R_{bBh}
+\frac{\operatorname{sd}_{b}(R_{bBh})}{\sqrt{N_\zeta}}.
\end{aligned}\label{eq:correction-block-main}\end{equation}
{Retaining the guide yields a score of zero by construction, so a negative $\mathcal S_{Bh}$ favors a proposal under the working risk criterion. Conditional on the fixed observations and estimated working covariance, the standard-error term measures Monte Carlo variation in score computation across random perturbations. The perturbation and block settings are specified in Supplementary Section~C.1.5, and the working covariance in Supplementary Section~C.1.4.}

{Using a policy selected on separate validation subjects and fixed thereafter, we either select the lowest negative-score candidate or combine up to four negative-score candidates. If none has a negative score, we set the block's geometric weights to zero, so that it contributes no geometric penalty to the joint objective.}

\FloatBarrier
\subsubsection{Joint recovery of the full trajectory}
{After retrieving windows and selecting constraints locally, we recover the full trajectory by optimizing jointly across overlapping windows. With $Z\in\mathbb R^{L\times J\times3}$ expressed in the same coordinates as $Y$, we solve}
\begin{equation}\begin{split}
\widehat Z=\operatorname*{arg\,min}_Z\ \mathcal L_{\rm obs}(Z,Y)
+\lambda_t\mathcal T(Z)\\
+\frac{\eta_c}{2}\sum_{g,h}\gamma_g\beta_{gh}c_{gh}
(E_gZ-\mu_{gh})^\top\Pi_{gh}(E_gZ-\mu_{gh}).
\end{split}\label{eq:correction-combine}\end{equation}
The first term fits the observations, the second regularizes time variation, and the third penalizes departures from the selected local structure. Here $\beta_{gh}$ is a selection or mixture weight, $\gamma_g$ accounts for window overlap, and $\eta_c$ controls overall geometric strength. {The observation and temporal terms are those defined in Section~\ref{sec:correction}, with the latter given by~\eqref{eq:correction-temporal}.}

Supplementary Sections~C.1.1 and~C.1.6 give the observation loss, overlap weights and solver.

\Needspace{25\baselineskip}
\begin{algorithm}{Self-Action CMAT}
\AI{Input}{Observed joint trajectory $Y^{\rm raw}$ and fixed correction settings.}
\AI{Output}{A corrected joint trajectory $\widehat Y$ of the same length.}
\AL{Save root positions $b_t$, estimate one body scale $s_{\rm body}$, and compute $Y_{t,j}=(Y^{\rm raw}_{t,j}-b_t)/s_{\rm body}$.}
\AL{Fit the guide using~\eqref{eq:correction-guide}; estimate working covariance and residual guide variance.}
\AL{For each nine-frame query, retrieve comparable windows in $G$, compute relative coordinates, fit preliminary directions and construct cylindrical support $\mathcal C_g$.}
\AL{Compute four neighborhood weight vectors. Form 16 candidates using~\eqref{eq:correction-center-main}--\eqref{eq:correction-projections-main}, and adjust constraint strength for residual variation.}
\AL{Apply paired small perturbations to the full observations and recompute guides, matches and candidates. Score against retaining the guide and aggregate within time blocks.}
\AL{Use the fixed policy to select the lowest negative-score candidate or combine up to four negative-score candidates; set block geometry weights to zero if none is negative.}
\AL{Adjust overlap weights, fix local constraints and solve~\eqref{eq:correction-combine} jointly. Restore body scale and root positions and return the result.}
\end{algorithm}

\FloatBarrier\begin{figure}[H]
\centering
\cmatgraphic{3}{\includegraphics[width=\linewidth]{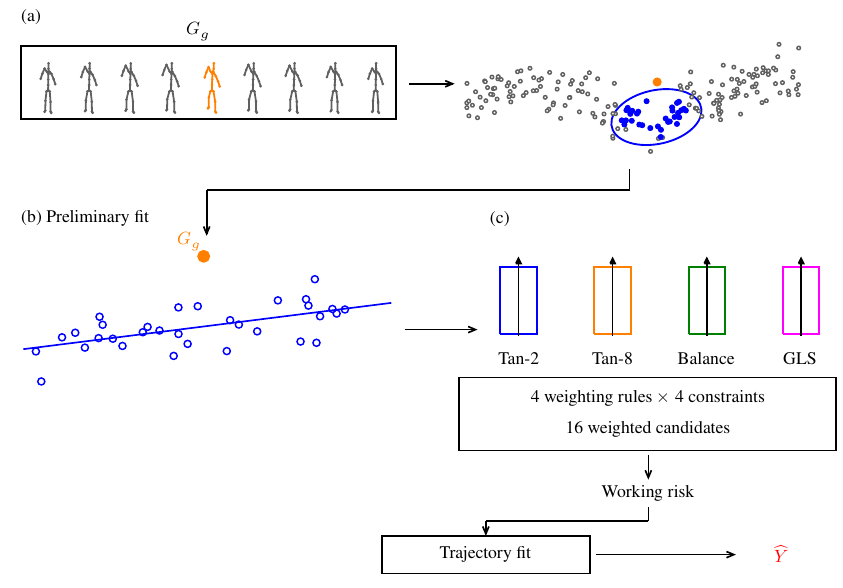}}
\caption{{Local geometry and trajectory recovery in Self-Action CMAT. (a) A temporal-guide window is matched to other windows in the same sequence; orange marks the query and blue the selected matches. (b) Preliminary local neighborhood structure. (c) Differently weighted candidate constraints are compared and combined with observation fit and temporal regularization for trajectory recovery.}}
\label{fig:method-3}
\end{figure}

\FloatBarrier
\FloatBarrier
\subsection{Rarity-preserving CMAT: generating poses along local structure}\label{sec:generation}
{To generate new poses in sparsely represented regions, we begin with a rare anchor, propose variations along locally fitted tangent directions and refine the resulting candidates. During candidate comparison, refinement and final screening, we enforce the same three feasibility conditions while evaluating local support against the neighborhood associated with each stage.}

\begin{samepage}
For query $q$ and neighbors $\mathcal L$, let $c_q$ be the center of the local fit in relative-rotation coordinates, with $q$ as the origin, and let $P_{N,q}$ be the normal projection from that fit. Define
\begin{equation}
E(q;\mathcal L)=\frac{\|P_{N,q}c_q\|}{\max\{\sqrt{\operatorname{median}_{r\in\mathcal L}\|\operatorname{Log}_q(r)\|^2},10^{-12}\}}.
\label{eq:chart-audit}
\end{equation}
The numerator is the distance from the query to the local affine fit; the denominator expresses it relative to neighborhood scale. We write the common feasibility conditions as
\end{samepage}
\begin{equation}
\mathsf F(q;\mathcal L):\quad
\rho_{g,-h}(q)\geq0.95,\qquad D_1(q)>\nu_g,\qquad E(q;\mathcal L)\leq\tau_E.
\label{eq:generation-feasible}
\end{equation}
{Here, for an anchor from subject $h$, $\rho_{g,-h}$ is the rarity percentile obtained by excluding all sequences of $h$ from the reference and recomputing both query and independent calibration scores against the reduced reference. The distance $D_1(q)=\min_{r\in\mathcal R}d(q,r)$ is measured against the complete reference, which is also used to retrieve local neighbors. These conditions require calibrated rarity, separation from stored poses and local geometric support. We set the separation threshold $\nu_g$ to the weighted 1st percentile of independent calibration poses' nearest complete-reference distances, floored at $10^{-6}$, and use the local-deviation threshold $\tau_E=0.55$.}

\FloatBarrier
\subsubsection{Exploring and comparing candidates}
{We first sample search poses from the reference and retain rare anchors with available partners from the same subject. In each attempt, we sample an anchor $q_a$ uniformly, select a different rare partner $q_p$ from that subject and retrieve the anchor neighbors $\mathcal L_a$ from the complete reference. From the local fit, we obtain an orthonormal basis $T_a$, projection $P_{T,a}=T_aT_a^\top$, neighborhood scale $\ell_a$ and leading singular values $\sigma_{a,1},\ldots,\sigma_{a,d_g}$ of the weighted centered matrix.}

{For the relative partner direction $e_a=\operatorname{Log}_{q_a}(q_p)$, we combine its tangent component with the observed neighborhood variation to form the initial proposals:}
\begin{equation}\begin{aligned}
v_b&=\tau_bP_{T,a}e_a+\kappa T_a\operatorname{diag}(\sigma_{a,1},\ldots,\sigma_{a,d_g})z_b,\\
q_b^{(0)}&=\operatorname{Exp}_{q_a}\{\mathcal B_a(v_b)\},\qquad z_b\sim N(0,I_{d_g}).
\end{aligned}\label{eq:generation-proposal}\end{equation}
Here $\mathcal B_a$ bounds the total increment relative to $\ell_a$ and caps each joint rotation. The fixed proposal settings are $B=6$ candidates per attempt, $\kappa=0.34$ with $\tau_b\sim\operatorname{Unif}(0.25,0.75)$ and $z_b\sim N(0,I_{d_g})$ drawn independently of each other and across candidates; pairing and capping details are given in Supplementary Sections~C.4.1--C.4.2.

{For each initial proposal $q_b^{(0)}$, we refit the local geometry using the fixed anchor neighbors, expressed in relative coordinates $\xi_i=\operatorname{Log}_{q_b^{(0)}}(q_i)\in\mathbb R^{63}$. With the fitted orthonormal basis $T$ and projections $P_T=TT^\top$ and $P_N=I-P_T$, we compute scores for four geometric heads as}
\begin{equation}
L_{ih}(q)=-a_h\frac{\|P_T\xi_i\|^2}{s_T^2}
-b_h\frac{\|P_N\xi_i\|^2}{s_N^2}
-q_h\frac{\|\xi_i\|^2}{s_D^2}.
\label{eq:attention-score-general}
\end{equation}
The scales $s_T^2,s_N^2,s_D^2$ are the medians of the corresponding squared distances, with positive numerical lower bounds.

{We average the four head displacements from~\eqref{eq:token-attention} to obtain $h$, then form the damped update}
\begin{equation}
\delta=g\{P_Nh+\eta P_Th\}.
\label{eq:cmat-update}
\end{equation}
{Here $P_Nh$ is the normal component of the four-head mean displacement, and $\eta P_Th$ is its tangential component scaled by $\eta$. The factor $g$ controls total step size and decreases when weights are overly concentrated; its calculation and the head coefficients are given in Supplementary Section~C.3. We obtain each candidate $q_b$ by applying the rotation update~\eqref{eq:pose-update} with joint and anchor-relative bounds.}

{To compare candidates against the same local observations, we keep $\mathcal L_a$ fixed and select the candidate with the smallest $E(q_b;\mathcal L_a)$ among those meeting~\eqref{eq:generation-feasible}. If none qualifies, we forward the first candidate to refinement for a further repair attempt, without accepting it at this stage.}

\FloatBarrier
\subsubsection{Refining the selected candidate}
{At the selected candidate $q_c$, we retrieve a fresh neighborhood $\mathcal L_c$ to estimate the current local structure, compute a CMAT increment using~\eqref{eq:cmat-update} and cap it jointwise to obtain $\delta_c$. We then evaluate update proportions $u$ from a fixed set $\mathcal G$ containing zero:}
\begin{equation}
q(u)=\operatorname{Exp}_{q_c}(u\delta_c),\qquad
\rho_{g,-h}\{q(u)\}\geq\rho_{g,-h}(q_c)-\varepsilon_R.
\label{eq:generation-refinement}
\end{equation}
{For each trial, we refit using the fixed $\mathcal L_c$ and recompute rarity and $D_1$, then choose as $q_f$ the trial with the smallest $E$ among those satisfying both~\eqref{eq:generation-feasible} and the rarity-loss bound in~\eqref{eq:generation-refinement}. If none qualifies, we retain $q_f=q_c$. The step sizes and capping rules for refinement are given in Supplementary Section~C.4.}

\FloatBarrier
\subsubsection{Final screening and yield}
{For final screening, we retrieve neighbors $\mathcal L_f$ anew at $q_f$ and re-evaluate~\eqref{eq:generation-feasible}, thereby assessing the output in its own neighborhood rather than against the anchor or refinement neighborhoods. We also require finite rotation parameters and SMPL joint coordinates, with normalized distances between designated joint pairs at least as large as their calibrated lower bounds. We accept $q_f$ only if all checks pass, including when refinement began from an infeasible candidate.}

{We accept at most one pose per attempt. With $I_m$ indicating acceptance in attempt $m$, we report the yield $\widehat p_{\rm acc}=M^{-1}\sum_{m=1}^M I_m$ as the fraction of all attempts producing an eligible pose. Detailed screening specifications are given in Supplementary Section~C.4.5.}

\Needspace{30\baselineskip}
\begin{algorithm}{Rarity-preserving CMAT}
\AI{Input}{Reference and independent calibration sets, fixed fitting and screening settings, step set $\mathcal G$, and attempt count $M$.}
\AI{Output}{Accepted poses $\mathcal A$ and yield $\widehat p_{\rm acc}$.}
\AL{Calibrate rarity and separation as in Section~\ref{sec:rarity} and~\eqref{eq:generation-feasible}; retain rare anchors with same-subject partners. Set $\mathcal A=\varnothing$.}
\AL{\textbf{For} $m=1,\ldots,M$: sample anchor $q_a$ and partner $q_p$; retrieve $\mathcal L_a$ and fit the anchor geometry.}
\AL[1]{Generate $B$ initial candidates by~\eqref{eq:generation-proposal}; refit using $\mathcal L_a$ and update each by~\eqref{eq:attention-score-general}--\eqref{eq:cmat-update}, with joint and anchor-relative bounds.}
\AL[1]{Among candidates satisfying~\eqref{eq:generation-feasible} on $\mathcal L_a$, choose the smallest-$E$ candidate $q_c$; if none qualifies, use the first.}
\AL[1]{Retrieve $\mathcal L_c$, refit and compute the joint-capped increment $\delta_c$. Form all trials in~\eqref{eq:generation-refinement} and refit them using $\mathcal L_c$.}
\AL[1]{Choose the smallest-$E$ trial satisfying~\eqref{eq:generation-feasible} and~\eqref{eq:generation-refinement} as $q_f$; if none qualifies, set $q_f=q_c$.}
\AL[1]{Retrieve $\mathcal L_f$; apply~\eqref{eq:generation-feasible}, finiteness and SMPL joint-pair checks. Add $q_f$ to $\mathcal A$ only if all checks pass.}
\AL{\textbf{End for}. Return $\mathcal A$ and $\widehat p_{\rm acc}=|\mathcal A|/M$.}
\end{algorithm}
\paperfigure{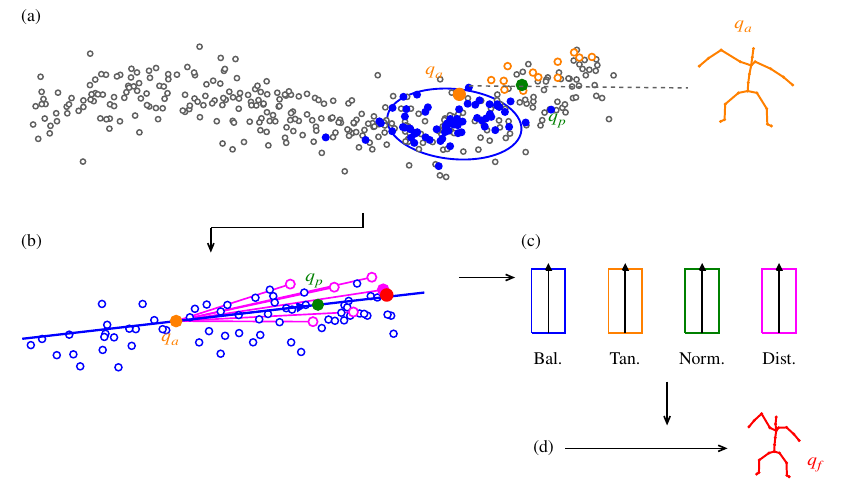}{{Rarity-preserving CMAT generation. (a) The orange anchor, green partner and blue geometric neighbors. (b) Candidate exploration within the fixed anchor neighborhood; magenta open circles are candidates and red marks the selected pose. (c) Neighbor retrieval and refinement at the selected candidate. (d) Renewed neighbor retrieval for final screening.}}

\FloatBarrier
\section{Experiments}\label{sec:experiments}
We first evaluate motion correction, then the sensitivity of rarity calibration to reference size, and finally rare-pose generation.

The motion data used in the experiments are drawn from AMASS, which integrates multiple motion-capture
datasets and is available at \url{https://amass.is.tue.mpg.de/}.
We use the SMPL body model to obtain joint
positions and body surfaces from pose and shape parameters;
the model is available at \url{https://smpl.is.tue.mpg.de/}.
For the sequence-correction comparisons, the official SmoothNet
implementation and pretrained models
are available through \url{https://github.com/cure-lab/SmoothNet}.

\FloatBarrier
\subsection{Controlled correction of rare motion}\label{sec:exp-correction}
{To evaluate correction of rare AMASS configurations in their temporal context, we first exclude subjects used for development, parameter selection or rarity calibration from the fixed reference of three million poses, leaving 2,454,257 eligible poses from 142 subjects. For each query subject, we then remove that subject from the reference and recompute independent source calibration against the reduced reference, as described in Section~\ref{sec:rarity}; applying $\rho_{g,-h}\geq0.95$ yields 145,732 rare poses from 123 subjects.}

{Using a fixed ordering of rare anchors, we select up to four clips per subject, prioritize different motions and prohibit overlap within a recording file; each clip spans 240 frames at 30 Hz. This procedure yields 236 clips from 99 subjects, comprising 56,640 clean context frames and 13,906 rare frames.}

We perturb the local rotations of the 21 body joints:
\begin{equation}\begin{aligned}
\widetilde R_{t,j}&=R_{t,j}\operatorname{Exp}(\epsilon_{t,j}),\\
\epsilon_{t,j}&=\sigma\{\sqrt{0.8}\,a_{t,j}+\sqrt{0.2}\,b_{t,j}\},\qquad \sigma=10\pi/180,\\
a_{t,j}&=\rho a_{t-1,j}+\sqrt{1-\rho^2}\,z_{t,j},\qquad \rho=\exp(-1/15).
\end{aligned}\label{eq:exp-noise}\end{equation}
The initial states, innovations and $b_{t,j}$ are independent standard three-dimensional Gaussians.

Artificial rotation noise can amplify motion variation and produce trajectories with excessive amplitude. To limit such changes, we use the empirical 90th percentile of clip amplitude to define an upper bound. Amplitude is the root-mean-square displacement of root-relative joint positions about their respective temporal means, divided by body scale. It is computed separately for the two noise realizations of each clip and then averaged. Rounding this percentile upward to a grid of 0.01 gives a bound of 0.73 body scales; clips with mean amplitude no greater than this bound are retained.

{Under this empirical amplitude bound, we retain 212 clips from 93 subjects, comprising 25,598 rare frame-by-realization observations; 25,510 of these have consecutive preceding and following frames and can therefore be evaluated using centered second differences.}

{The comparisons are temporal correction alone, the basic local estimator (YY-based), and SmoothNet.} The YY-based comparator and CMAT share normalization, temporal guidance, retrieved windows, four projection choices, displacement control, candidate-comparison budgets and the joint solver. {(The YY-based comparator uses uniform weights within cylinders with candidate support sizes, rather than the axial--transverse kernel-product weights of the original YY estimator.)} {We tune the geometric strength of each method separately under the same validation procedure and candidate-selection rules, giving both methods the same scope for parameter selection. Supplementary Section~A provides the configurations.}

Correction is evaluated by MPJPE, PA-MPJPE and acceleration error on rare frames. Let $\mathcal I$ index rare-frame observations, and let $p_{t,j}$ and $\widehat p_{t,j}$ be the root-relative target and corrected coordinates of joint $j$, with $J$ joints. MPJPE is the mean Euclidean distance between corresponding joints:
\begin{equation}
\operatorname{MPJPE}=\frac{1}{|\mathcal I|J}\sum_{t\in\mathcal I}\sum_{j=1}^{J}\|\widehat p_{t,j}-p_{t,j}\|.
\end{equation}
PA-MPJPE first fits an optimal similarity transform to all joints within each frame. Define
\begin{equation}
(s_t^*,R_t^*,b_t^*)\in\operatorname*{arg\,min}_{s\geq0,\,R\in\mathrm{SO}(3),\,b\in\mathbb R^3}
\sum_{j=1}^{J}\|sR\widehat p_{t,j}+b-p_{t,j}\|^2.
\end{equation}
The resulting mean joint distance is
\begin{equation}
\operatorname{PA\text{-}MPJPE}=\frac{1}{|\mathcal I|J}\sum_{t\in\mathcal I}\sum_{j=1}^{J}
\|s_t^*R_t^*\widehat p_{t,j}+b_t^*-p_{t,j}\|.
\end{equation}
Acceleration error compares second differences of consecutive frames. With $\Delta^2p_{t,j}=p_{t+1,j}-2p_{t,j}+p_{t-1,j}$ and $\mathcal I_{\rm acc}$ denoting rare centers with consecutive preceding and following frames, define
\begin{equation}
E_{\rm acc}=\frac{1}{|\mathcal I_{\rm acc}|J}\sum_{t\in\mathcal I_{\rm acc}}\sum_{j=1}^{J}
\|\Delta^2\widehat p_{t,j}-\Delta^2p_{t,j}\|.
\end{equation}
The spatial errors are reported in mm and acceleration error in mm/frame$^2$. {(When multiple clips and noise realizations are available, their valid frame observations are pooled before averaging as defined above.)}

\FloatBarrier
\Needspace{40\baselineskip}
\subsubsection{Examples of recovered motion}

{To illustrate recovery within a single sequence, Figure~\ref{fig:exp-6} traces framewise errors for CMU/18\_19\_rory under the first noise realization. We select this clip because its rare-frame PA-MPJPE reduction from perturbed input to CMAT, 19.608 mm, is closest to the retained-cohort median of 19.591 mm; the yellow shading marks its 50 rare frames within the 240-frame sequence.}

\expfigure{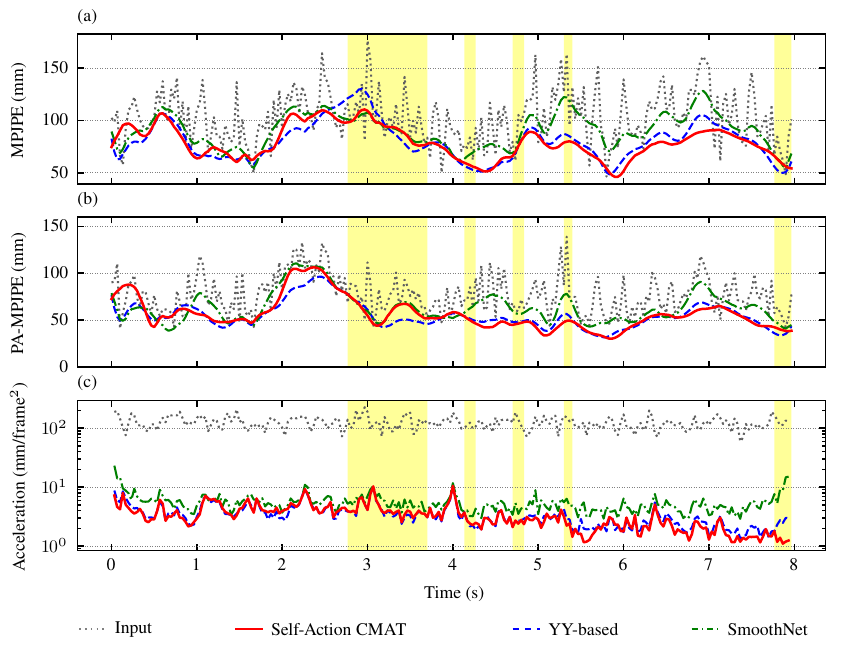}{{Framewise errors in the representative retained clip CMU/18\_19\_rory, using the first noise realization. Yellow bands mark its 50 rare frames.}}

\FloatBarrier
\Needspace{16\baselineskip}
\subsubsection{Comparison with geometric fitting and a pretrained model}
\leavevmode\par

\begin{table}[H]
\centering\small
\setlength{\tabcolsep}{4pt}
\caption{Rare-frame errors in the exploratory evaluation with amplitude at most 0.73 body scales. All methods use identical observations; lower is better.}\label{tab:bounded-main}
\begin{tabular}{lrrr}\hline
Method & MPJPE (mm) & PA-MPJPE (mm) & Acc. (mm/frame$^2$)\\\hline
Perturbed input & 108.227 & 83.228 & 129.440\\
Temporal only & 92.738 & 71.500 & 4.285\\
{YY-based} & 82.385 & 62.681 & 3.320\\
SmoothNet & 94.038 & 70.732 & 5.409\\
Self-Action CMAT & 81.075 & 61.990 & 3.146\\
\hline\end{tabular}
\end{table}

\Needspace{28\baselineskip}
{Within the selected evaluation population, CMAT has lower joint-position, aligned-pose and acceleration errors than the perturbed input, temporal correction alone, the YY-based comparator and pretrained SmoothNet (Table~\ref{tab:bounded-main}). Figure~\ref{fig:exp-5} reports comparator-minus-CMAT improvements with pointwise 95\% percentile intervals from 10,000 paired subject-bootstrap resamples. These intervals are conditional on the retained population and fixed empirical amplitude bound of 0.73 body scales; all lie above zero, supporting mean error reductions under these evaluation conditions.}

\expfigure{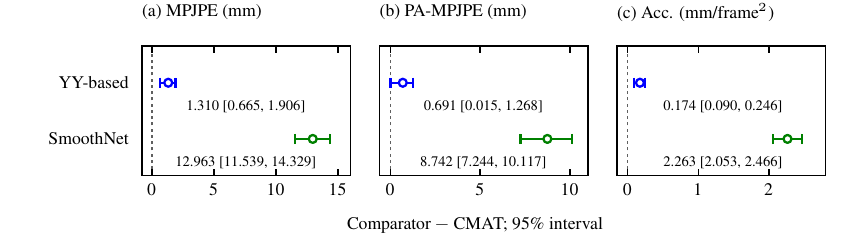}{{Correction comparisons for 212 clips from 93 subjects in the selected evaluation population.}}

{Because rarity screening leaves few frames in 3DPW--EFT, Supplementary Section~E uses all 35,515 frames to evaluate correction without additional noise; CMAT reduces PA-MPJPE from 52.194 to 50.611 mm.}

\FloatBarrier
\subsubsection{Effects of individual processing steps}

{To examine the contributions of local geometry, weighting and candidate selection, Table~\ref{tab:bounded-ablation} presents representative component analyses. Their pointwise 95\% intervals all lie above zero; complete results are provided in Supplementary Section~D.2.}

\begin{table}[!htbp]
\centering\small
\setlength{\tabcolsep}{4pt}
\caption{Selected component comparisons on the same selected population. PA-MPJPE is variant minus full CMAT; positive changes favor full CMAT. Intervals are pointwise 95\% intervals, conditional on the selected bounds.}\label{tab:bounded-ablation}
\begin{tabular}{lrr}\hline
Change & $\Delta$ PA (mm) & 95\% interval\\\hline
Temporal only & 9.510 & [7.858, 10.998]\\
{Reliability-only weights} & 0.872 & [0.634, 1.104]\\
Without GLS & 0.582 & [0.298, 0.867]\\
Equal candidate mixing & 1.655 & [0.866, 2.402]\\
Normal projection only & 6.047 & [4.981, 7.037]\\
Without final temporal term & 0.118 & [0.099, 0.135]\\
\hline\end{tabular}
\end{table}

\FloatBarrier
\subsection{Rarity stability across reference sizes}\label{sec:exp-stability}
{To assess whether a fixed reference can provide reproducible rarity classifications for generation, we construct a benchmark with three million AMASS reference poses and separate calibration and evaluation cohorts. With at most 16 uniformly sampled poses per sequence, the calibration cohort contains 2,272 poses from eight subjects, and the fixed evaluation cohort contains 3,302 poses from 207 sequences and nine different subjects.} 

{We vary the reference size across 100,000, 500,000, one million, two million, 2.5 million and three million poses. At each size below three million, we draw three reference subsets, recalibrate rarity separately for each subset and evaluate stability across all three pairwise comparisons.} 

{For pose $i$ in sequence $s$ of subject $h$, we assign weight $w_i=(H L_h m_{hs})^{-1}$, where $H$ denotes the number of evaluation subjects, $L_h$ the number of sequences for subject $h$, and $m_{hs}$ the number of evaluated poses in sequence $s$. This weighting gives equal contributions to subjects and, within each subject, to sequences. We summarize selection size by the weighted rare fraction and reproducibility by weighted overall label agreement and rare-set Jaccard overlap; unweighted Spearman correlation and median absolute percentile change measure ranking and numerical stability (Supplementary Section~B.4).}

\Needspace{32\baselineskip}
Figure~\ref{fig:exp-7} shows that larger references improve agreement of labels and rare sets across repeated draws while reducing changes in calibrated percentiles. Rarity ordering is already relatively stable for smaller references, yet the selected lists still change. Ranking agreement alone therefore does not establish reproducibility of rare-pose selection. Subsequent generation uses a fixed reference of three million poses and its calibration results, so anchor selection, candidate comparison and final screening share a consistent rarity standard.

\expfigure{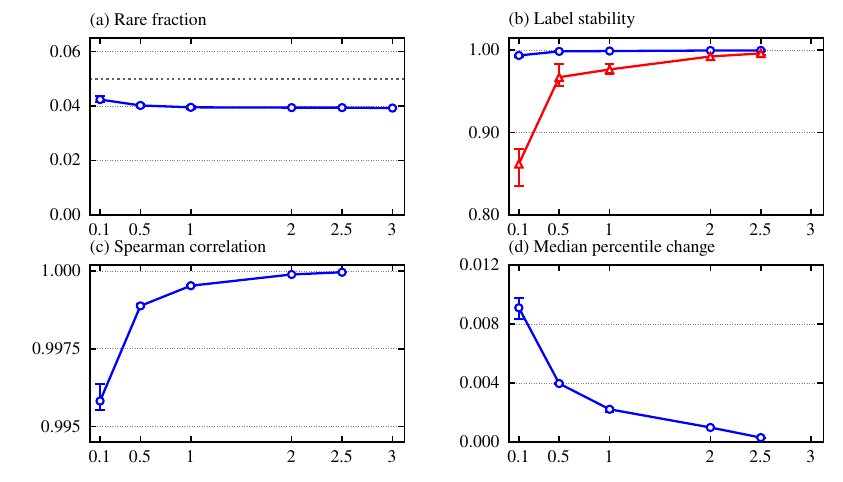}{{Effect of reference size on rarity assessment. Horizontal axes give reference poses in millions. (a) The horizontal dotted line marks 0.05. (b) Blue open circles denote label agreement and red open triangles the rare-set Jaccard coefficient.}}

\FloatBarrier
\subsection{Rare-pose generation}\label{sec:exp-generation}
\FloatBarrier
\subsubsection{Generation setup and acceptance criteria}
{We evaluate rare-pose generation using individual AMASS poses, with a fixed reference of three million poses and an independent calibration cohort of 2,272 poses. From 250,000 search poses, we retain 16,440 rare anchors satisfying $\rho_{g,-h}\geq0.95$ and having eligible same-subject partners.}

Using Rarity-preserving CMAT from Section 2, we generate candidates near rare anchors, compare and refine them while preserving rarity, and check their geometric support, rarity and difference from existing poses. We perform 50,000 generation attempts and define yield as the number of finally accepted poses divided by the total number of attempts.

\FloatBarrier
\subsubsection{Yield and screening results of the complete algorithm}
{Rarity-preserving CMAT produces 17,631 acceptable poses in 50,000 attempts, giving a yield of $35.262\%$.} Table~\ref{tab:generation-screen} reports individual screening rates: rarity and local fitting support reach $91.88\%$ and $99.97\%$, while non-memorization reaches $53.40\%$.

\begin{table}[!htbp]
\caption{Screening results for the complete Rarity-preserving CMAT algorithm.}
\label{tab:generation-screen}\centering\small
\begin{tabular}{lr}\hline
Criterion & Pass rate (\%)\\\hline
Rotation, finiteness and joint-pair distances & 71.48\\
Rarity & 91.88\\
Non-memorization & 53.40\\
Local fitting support & 99.97\\
All criteria jointly satisfied (final yield) & 35.262\\\hline
\end{tabular}
\end{table}

\FloatBarrier
\subsubsection{Comparison of pose generation approaches}
We consider five baseline generators to examine different ways of expanding rare-pose samples. All methods output poses in the SMPL body-joint rotation representation.

{\textbf{Rare replay.} This control outputs previously screened rare reference poses, testing whether direct duplication can satisfy the screening requirements for expanding the sample.}

{\textbf{Axis-angle Gaussian.} Composing existing poses with random joint-rotation increments produces nearby configurations, providing a test of unstructured perturbation.}

{\textbf{Part mix.} After partitioning joints into body parts, we recombine corresponding part rotations from different observed poses. The resulting full-body configurations preserve observed rotations within each part while changing the coordination across parts.}

{\textbf{PoseTrans-inspired 3D.} We adapt the tail-expansion strategy of PoseTrans to the shared SMPL joint-rotation representation, transforming existing rare poses.}

{\textbf{PoseAug-style 3D.} PoseAug's transformation-based augmentation approach motivates a second comparator in SMPL joint-rotation space.}

The five baselines and Rarity-preserving CMAT use the same three million reference poses and 16,440 rare anchors, with 50,000 attempts per method and the same acceptance criteria. Each attempt uses the same anchor across generators, together with its source, excluded subject and corresponding calibration context.  For every method, the non-memorization distance threshold is the weighted 1st percentile of distances from independent calibration poses to their nearest complete-reference poses, and the local deviation ratio must not exceed $0.55$. Table~\ref{tab:generation-precursor} and Figure~\ref{fig:exp-8} summarize results under this common evaluation protocol.

\begin{table}[!htbp]
\caption{Accepted counts and yields of generators under the common evaluation settings.}
\label{tab:generation-precursor}\centering\small
\begin{tabular}{lrrr}\hline
Method & Attempts & Accepted & Yield (\%)\\\hline
Rare replay & 50,000 & 0 & 0.000\\
Axis-angle Gaussian & 50,000 & 9,847 & 19.694\\
Part mix & 50,000 & 15,060 & 30.120\\
PoseTrans-inspired 3D & 50,000 & 11,978 & 23.956\\
PoseAug-style 3D & 50,000 & 9,625 & 19.250\\
Rarity-preserving CMAT & 50,000 & 17,631 & 35.262\\\hline
\end{tabular}
\end{table}

{Under the common evaluation protocol, Rarity-preserving CMAT has the highest yield at $35.262\%$ (17,631 accepted poses), followed by Part mix ($30.120\%$), PoseTrans-inspired 3D ($23.956\%$), Axis-angle Gaussian ($19.694\%$) and PoseAug-style 3D ($19.250\%$); Rare replay yields no accepted poses (Figure~\ref{fig:exp-8}(a)).}

\Needspace{28\baselineskip}
{The criterion-specific pass rates in Figure~\ref{fig:exp-8}(b) identify different screening limitations: replayed poses fail separation from the reference, while Gaussian perturbation and the two transformation baselines more often fail local fitting support. {CMAT maintains high pass rates for both rarity and local fitting support.}}

\expfigure{8}{{Generation yields and screening results. (a) Accepted poses per attempt (\%), with Wilson 95\% intervals. (b) Pass rates for the screening criteria (\%).}}

\FloatBarrier
\Needspace{34\baselineskip}
\subsubsection{Configurations of generated poses}
Figure~\ref{fig:exp-9} shows poses generated by the different methods and accepted after screening. Examples from the complete Rarity-preserving CMAT algorithm combine different torso inclinations, arm extensions and leg flexions, covering a range of full-body configurations.

\expfigure{9}{{Generated poses passing screening, sampled at fixed saved positions among accepted outputs. Replay yields no accepted poses because copied poses fail separation from the reference.}}

\FloatBarrier
\section{Discussion}\label{sec:discussion}
This work introduces attention into local manifold fitting so that the contribution of each neighboring observation can be adapted to the task. Correction and generation specify their own weights and output rules, applying the same local construction to trajectory recovery and the expansion of rare configurations. This design also leaves room for learned manifold fitting: attention scores or combinations of heads could be parameterized and learned from independent training data while retaining explicit neighborhood and geometric constraints.

{In single-sequence correction, retrieving many similar windows from one trajectory does not provide the same information as independent replicates: nearby windows share frames and estimation errors, and matches may lie asymmetrically around the local structure. Self-Action CMAT accounts for these features through distance weighting, center balancing and correlation-aware weights, followed by candidate comparison to determine the contribution of local geometry to trajectory recovery. The component comparisons support the roles of weighting and selection while indicating that improvements in spatial accuracy can involve a trade-off with temporal acceleration smoothness.}

{{Under the common screening criteria in this experiment, Rarity-preserving CMAT yields more accepted poses than the tested generators. The individual pass rates show that local fitting support is only one requirement: sufficient separation from stored poses also affects the final yield. These results show that generated poses pass the specified screens, but do not establish improvements in downstream models.}}

{Both applications demonstrate that manifold fitting without a separately trained network can support rare-motion recovery and constrained pose generation. In the controlled AMASS evaluation, Self-Action CMAT has lower errors than the pretrained SmoothNet comparator. These results support sequence-specific local geometry as an effective basis for correction without a pretrained correction network under the evaluated conditions. The two applications together demonstrate the feasibility of constructing outputs directly from available observations.}

{For correction, independently designed perturbations and new subjects can test the present findings; candidate selection under other error-dependence structures and constraints on bone length, contact and body geometry also merit study. For generation, accepted poses should be used to train downstream models under fixed training budgets and evaluated on independent rare-pose test sets. Continuous motion, mesh collisions and calibration across populations remain further directions. In addition, comparisons of untrained, partly learned and end-to-end learned approaches should include accuracy, latency, memory and the cost of adapting to new tasks, to establish when methods without task-specific training offer practical advantages.}

\FloatBarrier
\section*{Code availability}
The code for this study is available at
\url{https://github.com/zhigang-yao/CMAT/}.

\section*{Acknowledgments}

Zhigang Yao acknowledges support from the Singapore Ministry of Education through Tier 1 grants A8004146-00-00 and A-8002931-00-00 at the National University of Singapore. Feiyue Pan acknowledges support from the Shanghai Institute for Mathematics and Interdisciplinary Sciences (SIMIS). This work originated as a pilot project at the Interface of Data Science and Geometry (IDSG) Summer School held at SIMIS and was supported in part by SIMIS under grant SIMIS-ID-2025-OM. The authors also thank SIMIS for providing the resources and facilities essential to this work.


\section*{Supplementary material}
\noindent\textbf{Supplement to ``Local Manifold Attention for Recovery and Generation of Rare 3D Human Poses''}

\par\noindent{The separate file supp.pdf contains Sections A--F: experimental settings, evaluation metrics, implementation details, controlled comparisons, results for actual pose estimates, and Yao--Yau kernel formulas.}

\FloatBarrier
\bibliographystyle{cmat-plainnat}
\bibliography{references}

\begin{thebibliography}{17}
\providecommand{\natexlab}[1]{#1}
\providecommand{\url}[1]{\texttt{#1}}
\expandafter\ifx\csname urlstyle\endcsname\relax
  \providecommand{\doi}[1]{doi: #1}\else
  \providecommand{\doi}{doi: \begingroup \urlstyle{rm}\Url}\fi

\bibitem[Gong et~al.(2021)Gong, Zhang, and Feng]{gong2021poseaug}
Kehong Gong, Jianfeng Zhang, and Jiashi Feng.
\newblock {PoseAug}: A differentiable pose augmentation framework for {3D}
  human pose estimation.
\newblock In \emph{Proceedings of the {IEEE/CVF} Conference on Computer Vision
  and Pattern Recognition ({CVPR})}, pages 8575--8584, 2021.

\bibitem[He et~al.(2024)He, Tiwari, Birdal, Lenssen, and Pons-Moll]{he2024nrdf}
Yannan He, Garvita Tiwari, Tolga Birdal, Jan~Eric Lenssen, and Gerard
  Pons-Moll.
\newblock {NRDF}: Neural {Riemannian} distance fields for learning articulated
  pose priors.
\newblock In \emph{Proceedings of the {IEEE/CVF} Conference on Computer Vision
  and Pattern Recognition ({CVPR})}, pages 1661--1671, 2024.

\bibitem[Jiang et~al.(2022)Jiang, Jin, Liu, Qian, Luo, and
  Liu]{jiang2022posetrans}
Wentao Jiang, Sheng Jin, Wentao Liu, Chen Qian, Ping Luo, and Si~Liu.
\newblock {PoseTrans}: A simple yet effective pose transformation augmentation
  for human pose estimation.
\newblock In \emph{Computer Vision -- {ECCV} 2022}, volume 13665 of
  \emph{Lecture Notes in Computer Science}, pages 643--659, Cham, 2022.
  Springer.

\bibitem[Kocabas et~al.(2020)Kocabas, Athanasiou, and Black]{kocabas2020vibe}
Muhammed Kocabas, Nikos Athanasiou, and Michael~J. Black.
\newblock {VIBE}: Video inference for human body pose and shape estimation.
\newblock In \emph{Proceedings of the {IEEE/CVF} Conference on Computer Vision
  and Pattern Recognition ({CVPR})}, pages 5253--5263, 2020.

\bibitem[Li et~al.(2025)Li, Su, Lin, Yau, and Yao]{li2025metabolomic}
Bingjie Li, Jiaji Su, Runyu Lin, Shing-Tung Yau, and Zhigang Yao.
\newblock Manifold fitting reveals metabolomic heterogeneity and disease
  associations in {UK Biobank} populations.
\newblock \emph{Proceedings of the National Academy of Sciences}, 122\penalty0
  (22):\penalty0 e2500001122, 2025.

\bibitem[Loper et~al.(2015)Loper, Mahmood, Romero, Pons-Moll, and
  Black]{loper2015smpl}
Matthew Loper, Naureen Mahmood, Javier Romero, Gerard Pons-Moll, and Michael~J.
  Black.
\newblock {SMPL}: A skinned multi-person linear model.
\newblock \emph{ACM Transactions on Graphics}, 34\penalty0 (6):\penalty0
  248:1--248:16, 2015.

\bibitem[Mahmood et~al.(2019)Mahmood, Ghorbani, Troje, Pons-Moll, and
  Black]{mahmood2019amass}
Naureen Mahmood, Nima Ghorbani, Nikolaus~F. Troje, Gerard Pons-Moll, and
  Michael~J. Black.
\newblock {AMASS}: Archive of motion capture as surface shapes.
\newblock In \emph{Proceedings of the {IEEE/CVF} International Conference on
  Computer Vision ({ICCV})}, pages 5442--5451, 2019.

\bibitem[Pan and Yao(2026)]{cmat2026supp}
Feiyue Pan and Zhigang Yao.
\newblock Supplement to ``{Local Manifold Attention for Recovery and Generation
  of Rare 3D Human Poses}'', 2026.

\bibitem[Rempe et~al.(2021)Rempe, Birdal, Hertzmann, Yang, Sridhar, and
  Guibas]{rempe2021humor}
Davis Rempe, Tolga Birdal, Aaron Hertzmann, Jimei Yang, Srinath Sridhar, and
  Leonidas~J. Guibas.
\newblock {HuMoR}: {3D} human motion model for robust pose estimation.
\newblock In \emph{Proceedings of the {IEEE/CVF} International Conference on
  Computer Vision ({ICCV})}, pages 11488--11499, 2021.

\bibitem[Tiwari et~al.(2022)Tiwari, Anti{\'c}, Lenssen, Sarafianos, Tung, and
  Pons-Moll]{tiwari2022posendf}
Garvita Tiwari, Dimitrije Anti{\'c}, Jan~Eric Lenssen, Nikolaos Sarafianos,
  Tony Tung, and Gerard Pons-Moll.
\newblock {Pose-NDF}: Modeling human pose manifolds with neural distance
  fields.
\newblock In \emph{Computer Vision -- {ECCV} 2022}, volume 13665 of
  \emph{Lecture Notes in Computer Science}, pages 572--589, Cham, 2022.
  Springer.

\bibitem[Yao and Su(2026)]{yao2026review}
Zhigang Yao and Jiaji Su.
\newblock Manifold fitting: A review of methods and applications.
\newblock \emph{Harvard Data Science Review}, 8\penalty0 (3), 2026.

\bibitem[Yao and Xia(2025)]{yao2025manifold}
Zhigang Yao and Yuqing Xia.
\newblock Manifold fitting under unbounded noise.
\newblock \emph{Journal of Machine Learning Research}, 26\penalty0
  (45):\penalty0 1--55, 2025.

\bibitem[Yao et~al.(2023)Yao, Su, Li, and Yau]{yao2023manifold}
Zhigang Yao, Jiaji Su, Bingjie Li, and Shing-Tung Yau.
\newblock Manifold fitting.
\newblock Preprint, arXiv:2304.07680, 2023.

\bibitem[Yao et~al.(2024{\natexlab{a}})Yao, Li, Lu, and Yau]{yao2024singlecell}
Zhigang Yao, Bingjie Li, Yukun Lu, and Shing-Tung Yau.
\newblock Single-cell analysis via manifold fitting: A framework for {RNA}
  clustering and beyond.
\newblock \emph{Proceedings of the National Academy of Sciences}, 121\penalty0
  (37):\penalty0 e2400002121, 2024{\natexlab{a}}.

\bibitem[Yao et~al.(2024{\natexlab{b}})Yao, Su, and Yau]{yao2024cyclegan}
Zhigang Yao, Jiaji Su, and Shing-Tung Yau.
\newblock Manifold fitting with {CycleGAN}.
\newblock \emph{Proceedings of the National Academy of Sciences}, 121\penalty0
  (5):\penalty0 e2311436121, 2024{\natexlab{b}}.

\bibitem[Zeng et~al.(2022)Zeng, Yang, Ju, Li, Wang, and Xu]{zeng2022smoothnet}
Ailing Zeng, Lei Yang, Xuan Ju, Jiefeng Li, Jianyi Wang, and Qiang Xu.
\newblock {SmoothNet}: A plug-and-play network for refining human poses in
  videos.
\newblock In \emph{Computer Vision -- {ECCV} 2022}, volume 13665 of
  \emph{Lecture Notes in Computer Science}, pages 625--642, Cham, 2022.
  Springer.

\bibitem[Zhang et~al.(2024)Zhang, Bhatnagar, Xu, Winkler, Kadlecek, Tang, and
  Bogo]{zhang2024rohm}
Siwei Zhang, Bharat~Lal Bhatnagar, Yuanlu Xu, Alexander Winkler, Petr Kadlecek,
  Siyu Tang, and Federica Bogo.
\newblock {RoHM}: Robust human motion reconstruction via diffusion.
\newblock In \emph{Proceedings of the {IEEE/CVF} Conference on Computer Vision
  and Pattern Recognition ({CVPR})}, pages 14606--14617, 2024.

\end{thebibliography}


\begin{thebibliography}{1}
\providecommand{\natexlab}[1]{#1}
\providecommand{\url}[1]{\texttt{#1}}
\expandafter\ifx\csname urlstyle\endcsname\relax
  \providecommand{\doi}[1]{doi: #1}\else
  \providecommand{\doi}{doi: \begingroup \urlstyle{rm}\Url}\fi

\bibitem[Benjamini and Hochberg(1995)]{benjamini1995fdr}
Yoav Benjamini and Yosef Hochberg.
\newblock Controlling the false discovery rate: A practical and powerful
  approach to multiple testing.
\newblock \emph{Journal of the Royal Statistical Society, Series B},
  57\penalty0 (1):\penalty0 289--300, 1995.

\end{thebibliography}
\end{document}


\renewcommand{\thepage}{S\arabic{page}}
\title{Supplement to ``Local Manifold Attention for Recovery and Generation of Rare 3D Human Poses''}
\author{%
Feiyue Pan\textsuperscript{1}\quad
Zhigang Yao\textsuperscript{2,*}\\[0.8em]
\begin{minipage}{0.96\textwidth}\small\centering
\textsuperscript{1}School of Mathematical Sciences, Fudan University;\\
Shanghai Institute for Mathematics and Interdisciplinary Sciences,\\
Shanghai, China.\\[0.4em]
\textsuperscript{2}Department of Statistics and Data Science,\\
National University of Singapore, Singapore\\[0.4em]
\textsuperscript{*}Corresponding author:
\href{mailto:zhigang.yao@nus.edu.sg}{zhigang.yao@nus.edu.sg}
\end{minipage}}
\date{}
\maketitle

This supplement provides the settings, computational details and additional analyses for the main article. Sections A--F retain the original appendix numbering. References explicitly marked ``main article'' refer to that separate document.
\appendix
\setcounter{table}{0}
\renewcommand{\thetable}{\thesection.\arabic{table}}
\makeatletter\@addtoreset{table}{section}\makeatother

\section{Experimental parameters and reproducibility}\label{app:settings}

\begin{table}[!htbp]\centering\small
\caption{Settings of the two applications.}
\begin{tabular}{lll}\hline
Item & Correction & Generation\\\hline
Object dimension & $27J$ & 63\\
Local fitting rank & 3 & 12\\
Comparison objects & 32 windows & 64 poses\\
{Minimum support (CMAT)} & {$\min(24,K)$} & {$\min(14,K)$}\\
Maximum gate; tangent fraction & --- & 0.70; 0.10\\
Joint step cap & --- & $12^\circ$\\
{Candidate type/count} & {16 constraints} & {6 poses}\\
Refinement fractions & --- & $0,0.25,0.5,0.75,1$\\
Rarity decrease tolerance & --- & 0.005\\
Final local discrepancy bound & --- & 0.55\\
Rarity threshold (sampling) & $\rho_{g,-h}\geq0.95$ & $\rho_{g,-h}\geq0.95$\\
Upper-percentile rejection & none & none\\
Noise repetitions / attempts & 2 repetitions & 50,000 per method\\\hline
\end{tabular}\end{table}

\begin{table}[!htbp]\centering\small
\caption{Fixed deterministic head coefficients.}\label{tab:heads}
\begin{tabular}{lrrr}\hline
Head preference & $a_h$ & $b_h$ & $q_h$\\\hline
Balanced & 1 & 1 & 0.25\\
Closer in tangent coordinates & 4 & 1 & 0.25\\
Closer in normal coordinates & 1 & 4 & 0.25\\
Closer in total distance & 0.25 & 0.25 & 4\\\hline
\end{tabular}
\end{table}

\subsection{Correction parameters and comparison methods}
{In the controlled AMASS experiments, we decode both clean and perturbed rotations into 22 joint positions using the same neutral-shape SMPL proxy.}

{With $\lambda_t=12$ and $\tau_c=15$ frames held fixed, we compare $f\in\{0,0.8\}$, $\eta_c\in\{0.3,3,30,300\}$, and best-candidate or up-to-four-candidate mixing policies on the validation set. Crossing four weighting rules with four projections gives the 16 constraints described in Appendix~\ref{app:correction-local-geometry}.}

\section{Evaluation metrics and interval calculation}\label{app:metrics}
\subsection{Joint-position, aligned-pose and temporal errors}
{We calculate MPJPE as the mean Euclidean distance between target and estimated joints after separately centering both poses at the root. For PA-MPJPE, we first align the estimate to the target in each frame by solving}
\begin{equation}
(s_t^*,R_t^*,b_t^*)\in\operatorname*{arg\,min}_{s\geq0,\,R\in\mathrm{SO}(3),\,b\in\mathbb R^3}
\sum_j\|sR\widehat p_{t,j}+b-p_{t,j}\|^2,
\end{equation}

{and then average the Euclidean joint errors after applying the fitted similarity transformation. To quantify temporal error, we define $a_{t,j}=p_{t+1,j}-2p_{t,j}+p_{t-1,j}$ and its estimated counterpart $\widehat a_{t,j}$, and average their differences over the valid target centers $\mathcal I$:}
\begin{equation}
E_{\rm acc}=\frac1{|\mathcal I|J}\sum_{t\in\mathcal I}\sum_j
\|\widehat a_{t,j}-a_{t,j}\|.
\end{equation}

\subsection{Averaging errors and comparing methods}
If subject $h$ contributes $n_h$ valid frame observations and total error $S_h$, the frame-weighted error is
\begin{equation}
\widehat M_F=\frac{\sum_h S_h}{\sum_h n_h}.
\end{equation}

{For the controlled comparisons and component analyses, we construct intervals from 10,000 paired subject-level bootstrap resamples, conditional on the fixed reference, calibration cohort, clips, noise realizations and selection masks. Across the 19 component variants and three metrics, we compute two-sided normal-approximation $p$ values using the bootstrap standard deviation as the standard error and apply Benjamini--Hochberg adjustment~\citep{benjamini1995fdr} over the $m=57$ comparisons. The sorted adjusted values are $q_{(i)}=\min\{1,\min_{j\geq i}mp_{(j)}/j\}$, with the full component results reported in Appendix~\ref{app:controlled-supp}.}

\subsection{Yield and Wilson intervals}\label{app:generation-yield}
{For $M$ generation attempts, let $I_m$ indicate whether the output of attempt $m$ passes all acceptance conditions ($I_m=1$ for acceptance and $I_m=0$ otherwise). We calculate the accepted count and yield as}
\begin{equation}
A=\sum_{m=1}^{M}I_m,\qquad \widehat p_{\rm acc}=\frac{A}{M}.
\label{eq:yield}\end{equation}
{For the yields in Figure~9 of the main article, we report Wilson 95\% intervals with $z=1.96$ and endpoints}
\begin{equation}
\frac{\widehat p_{\rm acc}+z^2/(2M)\ \pm\ z\sqrt{\widehat p_{\rm acc}(1-\widehat p_{\rm acc})/M+z^2/(4M^2)}}{1+z^2/M}.
\end{equation}

\subsection{Rarity-stability metrics}\label{app:stability-metrics}
{For the fixed evaluation poses $q_1,\ldots,q_N$, we calculate the calibrated percentiles $\rho_i^{(a)}$ and rare labels $z_i^{(a)}=\mathbf1\{\rho_i^{(a)}\geq0.95\}$ under reference draw $a$. To weight subjects and their sequences equally, a pose $i$ in sequence $s$ of subject $h$ receives weight $w_i=(H L_h m_{hs})^{-1}$, where $H$ is the number of evaluation subjects, $L_h$ the number of sequences for that subject, and $m_{hs}$ the number of evaluated poses in that sequence; these weights satisfy $\sum_iw_i=1$.}

The rare fraction $P_a$, pairwise label agreement $A_{ab}$ and rare-set Jaccard coefficient $J_{ab}$ are
\begin{equation}\begin{aligned}
P_a&=\sum_{i=1}^{N}w_i z_i^{(a)},\\
A_{ab}&=\sum_{i=1}^{N}w_i\mathbf1\{z_i^{(a)}=z_i^{(b)}\},\\
J_{ab}&=\frac{\sum_iw_i z_i^{(a)}z_i^{(b)}}{\sum_iw_i\max\{z_i^{(a)},z_i^{(b)}\}}.
\end{aligned}\label{eq:stability-labels}\end{equation}
Here $P_a$ measures the classified rare fraction, $A_{ab}$ measures agreement over all labels, and $J_{ab}$ measures overlap between the rare lists.

For ranking and numerical stability, let $r_i^{(a)}$ be the rank of $\rho_i^{(a)}$ among the $N$ poses, using average ranks for ties, and write $\bar r^{(a)}=N^{-1}\sum_i r_i^{(a)}$. Spearman correlation and median absolute percentile change are
\begin{equation}\begin{aligned}
S_{ab}&=\frac{\sum_i(r_i^{(a)}-\bar r^{(a)})(r_i^{(b)}-\bar r^{(b)})}
{\sqrt{\sum_i(r_i^{(a)}-\bar r^{(a)})^2\sum_i(r_i^{(b)}-\bar r^{(b)})^2}},\\
D_{ab}&=\operatorname{median}_{1\leq i\leq N}|\rho_i^{(a)}-\rho_i^{(b)}|.
\end{aligned}\label{eq:stability-ranks}\end{equation}

\section{Implementation of correction and generation}\label{app:implementation}

\subsection{Implementing Self-Action CMAT}\label{app:correction-details}

\subsubsection{{Normalize the observations and compute the guide}}\label{app:correction-guidance}
{{For input $Y^{\rm raw}$, use root position $b_t$ (joint 0 for AMASS and the midpoint of hip joints 2 and 3 for the 3DPW--EFT data below). Normalize by}}
\begin{equation}\begin{aligned}
s_{\rm body}&=\max\left\{\operatorname{median}_t\left(\frac1J\sum_j
\|Y^{\rm raw}_{t,j}-b_t\|^2\right)^{1/2},10^{-8}\right\},\\
Y_{t,j}&=(Y^{\rm raw}_{t,j}-b_t)/s_{\rm body}.
\end{aligned}\end{equation}
{Holding this scale fixed over the sequence, we estimate a local variation scale for each joint from the third differences of the observations:}
\begin{equation}
s_j=\max\left\{\frac{\operatorname{median}_t\|(D^3Y)_{t,j}\|}{1.538\sqrt{20}},2\times10^{-5}\right\}.
\end{equation}
{Here $(D^3Y)_{t,j}=Y_{t+3,j}-3Y_{t+2,j}+3Y_{t+1,j}-Y_{t,j}$.}

{We define the residual weights and their associated loss by}
\begin{equation}
\omega_j(r)=\max\{0.02,\min(1,2.8s_j/\max(r,10^{-12}))\},\qquad
\psi_j(r)=\int_0^r u\omega_j(u)\,du.
\end{equation}
{We obtain the guide $G$ by minimizing $\mathcal L_{\rm obs}+12\mathcal T$, where $\mathcal L_{\rm obs}(Z,Y)=\sum_{t,j}\psi_j(\|Z_{t,j}-Y_{t,j}\|)$ and $\mathcal T$ is defined in~{\textup{(2.16)}} of the main article. The final residual weights $w_{t,j}=\omega_j(\|G_{t,j}-Y_{t,j}\|)$ are averaged over each matched window $i$ to obtain its reliability $r_i$.}

\subsubsection{{Comparable-window retrieval and support restriction}}

{We partition each sequence into overlapping 120-frame blocks with a 60-frame stride, adding a final block aligned with the last frame. Within each block, all contiguous nine-frame windows are eligible comparisons, and we query every fourth window along with the final window.}

{Writing $G_i$ for window $i$ in the block, $D$ for the within-window first difference and $t_i$ for the center frame, we standardize the retrieval-feature coordinates as follows:}
\begin{equation}\begin{aligned}
F_i&=\bigl(\operatorname{vec}G_i,\;0.70\operatorname{vec}(DG_i),\;
0.175\operatorname{vec}(D^2G_i)\bigr),\\
m_k&=\operatorname{median}_{i}F_{ik},\qquad
a_k=\operatorname{median}_{i}|F_{ik}-m_k|,\\
\widetilde F_{ik}&=\frac{F_{ik}-m_k}{\max\{a_k,10^{-3}\}},\qquad
\Delta_{ig}=\|\widetilde F_i-\widetilde F_g\|_2.
\end{aligned}\label{eq:retrieval-features}\end{equation}
{In~\eqref{eq:retrieval-features}, the premultiplication coefficients $0.70$ and $0.175$ affect $\Delta_{ig}$ only for coordinates whose MAD is below the numerical floor $10^{-3}$; otherwise they cancel under coordinate-wise standardization.}
Let $\operatorname{first}_m$ keep at most the first $m$ entries, $\operatorname{uniq}$ keep first occurrences, and $\Vert$ concatenate ordered lists:
\begin{equation}\begin{aligned}
\mathcal L_d&=\operatorname{sort}_{\Delta_{ig}}\{i:i\ne g\},\\
\mathcal L_{16}&=\operatorname{first}_{16}
\bigl(\operatorname{sort}_{\Delta_{ig}}\{i:|t_i-t_g|>9\}\bigr),\\
\mathcal L_t&=\operatorname{sort}_{|t_i-t_g|}\{i:0<|t_i-t_g|\leq18\},\\
\mathcal M_g&=\operatorname{first}_{32}
\bigl(\operatorname{uniq}(\mathcal L_{16}\Vert\mathcal L_t\Vert\mathcal L_d)\bigr).
\end{aligned}\label{eq:retrieval-matches}\end{equation}
{Using all $K$ retrieved position differences, we obtain the preliminary basis and cylinder by the construction in Appendix~\ref{app:cylinder-details}:}
\begin{equation}\begin{aligned}
x_i&=E_iG-G_g\in\mathbb R^{27J},\quad i\in\mathcal M_g,\qquad K=|\mathcal M_g|,\\
T_0&\in\mathbb R^{27J\times3},\quad T_0^\top T_0=I_3,\qquad
|\mathcal C_g|\geq\min(24,K),\\
\ell^2&=\max\{\operatorname{median}_{i\in\mathcal M_g}\|x_i\|^2,10^{-16}\},\\
s_T^2&=\max\{\operatorname{median}_{i\in\mathcal M_g}\|T_0^\top x_i\|^2,10^{-8}\ell^2\}.
\end{aligned}\label{eq:correction-fit-scales}\end{equation}

\subsubsection{{Four weighting rules and sixteen geometric candidates}}\label{app:correction-local-geometry}

{For the matched guide windows $G_g+x_i$, we express their offsets in the three preliminary directions of $T_0$ as $u_i=T_0^\top x_i$. All weights below are nonnegative and normalized on the common cylinder.}

\par\smallskip\noindent\textit{{Distance weights.}}\quad
{Two tangent-distance scales incorporate window reliability $r_i$:}
\begin{equation}
p_i^{(\tau)}\ \propto\ \mathbf1\{i\in\mathcal C\}\,r_i
\exp\left(-\frac{\|u_i\|^2}{\tau s_T^2}\right),\qquad \tau\in\{2,8\}.
\label{eq:correction-prior}
\end{equation}
{After normalization, the rule with $\tau=8$ has a broader distance decay than the rule with $\tau=2$.}

\par\smallskip\noindent\textit{{Balancing a one-sided neighborhood.}}\quad
{Starting from $p=p^{(8)}$, we calculate balance weights by penalizing a one-sided tangent center:}
\begin{equation}
\min_{w\geq0,\,\boldsymbol1^\top w=1}
\sum_iw_i\log\frac{w_i}{p_i}
+\frac{\kappa}{2}\left\|\sum_iw_i\bar u_i\right\|^2,
\qquad \kappa=\left(\sum_ip_i^2\right)^{-1}.
\label{eq:correction-balance}
\end{equation}
{The optimization is restricted to indices with $p_i>0$, and the rescaled coordinates $\bar u_i$ keep the query at the origin. Because balance is imposed through a soft penalty, we can also use neighborhoods for which exact balancing is infeasible.}

\par\smallskip\noindent\textit{{Accounting for correlated matches.}}\quad
{To account for shared errors in the GLS weights, we model correlation through the center-frame offsets $o_i$:}
\begin{equation}\begin{aligned}
C_{ik}&=(1-f)\mathbf1\{i=k\}+f\exp(-|o_i-o_k|/\tau_c),\\
A&=\operatorname{diag}(p^{-1/2})C\operatorname{diag}(p^{-1/2}),\\
w^{\rm GLS}&\in\operatorname*{arg\,min}_{w\geq0,\,\boldsymbol1^\top w=1}w^\top Aw.
\end{aligned}\label{eq:correction-gls}\end{equation}
{We solve this problem on the positive support of $p$, recovering $p$ when $f=0$. For positive $f$, the objective penalizes the joint weighting of correlated matches while retaining the distance and reliability factors.}

{\par\medskip\noindent Once the four weight vectors are fixed, we construct the corresponding correction candidates by refitting the center $\mu$ and orthonormal directions $b_1,b_2,b_3$ under each rule and forming}
\begin{equation}
\Pi^{(r)}=I-\sum_{k=1}^{r}b_kb_k^\top\quad(r=1,2,3),\qquad
\Pi^{(0)}=I.
\label{eq:correction-projections}
\end{equation}
{Each resulting constraint leaves $r$ tangent directions free, while $r=0$ constrains the full center; combining the four projections with the four weighting rules yields 16 candidates. For $d_{gh}=\Pi_{gh}(\mu_{gh}-G_g)$, the displacement coefficient and proposal are}
\begin{equation}
c_{gh}=\frac{v}{v+\|d_{gh}\|^2/D},\qquad
q_{gh}=G_g+\frac{\eta_c c_{gh}}{1+\eta_c c_{gh}}d_{gh},
\label{eq:correction-cap}
\end{equation}
{Here $D=27J$ and $v$ is the residual-guide variance described in Appendix~\ref{app:correction-working-variance}, with $c_{gh}=0$ when the denominator vanishes. After scoring $q_{gh}$, we use the selected center, projection and coefficient in the joint objective~{\textup{(2.22)}} of the main article.}

{All four weighting rules use the same preliminary basis and cylinder. We calculate numerically stable distance weights by subtracting the largest supported score before exponentiation, then multiplying by $r_i$ and normalizing. For the balance head, we scale each tangent coordinate by its weighted root mean square:}
\begin{equation}
a_k=\left(\sum_i p_i u_{ik}^2\right)^{1/2},\qquad
\bar u_{ik}=\frac{u_{ik}}{\max\{a_k,10^{-8}\max_l a_l,10^{-14}\}}.
\end{equation}
{Without subtracting a mean, we keep the query at the origin and solve~\eqref{eq:correction-balance} through the three-variable dual}
\begin{equation}
\min_b\ \log\left\{\sum_i p_i\exp(-b^\top\bar u_i)\right\}
+\frac{\|b\|^2}{2\kappa},
\end{equation}
{using Newton iterations with backtracking, and recover the weights as $w_i\propto p_i\exp(-b^\top\bar u_i)$. For the GLS simplex quadratic program, we use an active-set method, flooring positive entries of $p$ at $10^{-14}$ and renormalizing before calculating $p^{-1/2}$. The correlation time scale is fixed at $\tau_c=15$ frames.}

{To refit the geometry under each weight vector, we calculate $\bar x=\sum_iw_ix_i$ and the first three right singular vectors of $[\sqrt{w_i}(x_i-\bar x)^\top]_i$, giving the center $\mu=G_g+\bar x$. Storing the directions as rows of $B$ allows us to apply the four projections without forming a dense $D\times D$ matrix:}
\begin{equation}
\Pi=I-B^\top\operatorname{diag}(1-a)B,\qquad
a\in\{(0,1,1),(0,0,1),(0,0,0),(1,1,1)\}.
\end{equation}

\subsubsection{{Dependence and variance estimation for candidate scoring}}\label{app:correction-working-variance}
{
From the observations, we estimate both a covariance for perturbing the inputs during candidate comparison and a scalar $v$ describing the variation remaining after guiding. For the covariance, we vectorize each frame and take third differences, subtract each coordinate's temporal median, and estimate its robust scale as the MAD divided by $\Phi^{-1}(0.75)\approx0.6745$, the standard normal 75th percentile. We then clip deviations at four times this scale in either direction, using the root mean square if the MAD is zero but variation remains. After subtracting the clipped sample mean, we form the sample covariance $S_3$ and stabilize it by}
\begin{equation}
\widetilde S_3=\tfrac12 S_3+\tfrac12\operatorname{diag}(S_3),
\end{equation}
{and replace any numerically negative eigenvalues by zero.}

{We model temporal errors as a mixture of independent errors and an AR(1) process with $\rho_c=\exp(-1/\tau_c)$, giving lag covariance $k_f(d)=(1-f)\mathbf1\{d=0\}+f\rho_c^{|d|}$. To recover the position-scale covariance from the third differences, we divide by the variance of the same filter under this model, which for coefficients $a=(-1,3,-3,1)$ is}
\begin{equation}
V_3(f)=\sum_{r=0}^3\sum_{s=0}^3a_ra_s k_f(r-s).
\end{equation}
Let $P$ subtract the root position in each frame. Applying it also to the covariance gives $\Sigma_x=P\widetilde S_3P^\top/V_3(f)$ and preserves the root-centering constraint in simulated perturbations. The full-sequence working covariance is
\begin{equation}
\Sigma=K_f\otimes\Sigma_x,\qquad (K_f)_{ts}=k_f(t-s).
\end{equation}
{We hold the validation-selected fraction $f$ fixed when processing perturbed sequences, while recomputing the spatial covariance for each sequence.}

{To account for the modeled variation removed by smoothing, we approximate the guide's frequency response using its final residual and third-difference weights:}
\begin{equation}\begin{aligned}
v_{t,j}^{G}&=\{\|(D^3G)_{t,j}\|^2+(0.05s_j)^2\}^{-1/2},\\
\lambda_j&=\lambda_t s_j\operatorname{median}_t(v_{t,j}^{G})\big/\operatorname{mean}_t(w_{t,j}),\\
H_j(\omega)&=\{1+\lambda_j(2\sin(\omega/2))^6\}^{-1}.
\end{aligned}\end{equation}
The factor $H_j^2$ attenuates the temporal error spectrum. Integrating the retained spectrum and averaging over joints gives
\begin{equation}\begin{aligned}
S_f(\omega)&=(1-f)+f\frac{1-\rho_c^2}{1+\rho_c^2-2\rho_c\cos\omega},\\
v&=\frac1J\sum_{j=1}^J\bar\sigma_j^2\,
\frac1\pi\int_0^\pi H_j(\omega)^2S_f(\omega)\,d\omega,
\end{aligned}\end{equation}
{where $\bar\sigma_j^2$ is the mean of the three spatial-covariance diagonal entries for joint $j$. We evaluate the integral by the trapezoidal rule at 1,025 equally spaced frequencies and use the resulting sequence-level $v$ in the displacement coefficient in~\eqref{eq:correction-cap}.}

\subsubsection{Numerical settings for candidate scoring}\label{app:correction-selection}
{When evaluating main-paper equations~{\textup{(2.20)}}--{\textup{(2.21)}}, we set $N_\zeta=2$ and $\varepsilon=0.1$, initialize the AR(1) perturbations at stationarity, and share probes across candidates and methods. We group queries by their center frames into nonoverlapping 60-frame selection blocks and use $N_\zeta-1$ as the denominator in the sample variance underlying the sample standard deviation. For mixtures, the selected values $-\mathcal S_{Bh}$ are normalized to obtain $\beta_{gh}$.}

\subsubsection{Overlap weights and joint solution}\label{app:correction-joint-solver}
{To limit the cumulative constraint on a frame $t$ covered by $n_t$ query windows, we assign each query the overlap weight $\gamma_g=\min_{t\in g}n_t^{-1}$, with zero weight for unselected candidates.}

{Holding all $\mu_{gh},\Pi_{gh},c_{gh},\beta_{gh}$ fixed, we solve the main-paper objective~{\textup{(2.22)}} from the initial trajectory $Z=Y$ by iteratively reweighted least squares. At each iteration, we update the observation weights $\omega_j(\|Z_{t,j}-Y_{t,j}\|)$ and third-difference weights from the current trajectory, then solve the resulting linear system by preconditioned conjugate gradients. We allow at most 30 reweighting iterations and 800 iterations per linear solve, with relative tolerance $10^{-5}$.}

{Finally, we restore the saved scale and root positions to obtain $s_{\rm body}\widehat Z_{t,j}+b_t$.}

\subsection{Constructing the local fit and cylindrical support}\label{app:cylinder-details}

\subsubsection{Local directions for geometric scores and scale estimation}\label{app:local-geometry}
{With the query centered at the origin, let $\xi_i\in\mathbb R^p$ denote the $K$ retrieved comparison objects and $\ell^2=\max\{\operatorname{median}_i\|\xi_i\|^2,10^{-16}\}$ the squared neighborhood scale. We calculate the preliminary weights and center as}
\begin{equation}
\pi_i=\frac{\exp(-\|\xi_i\|^2/\ell^2)}{\sum_j\exp(-\|\xi_j\|^2/\ell^2)},\qquad c=\sum_i\pi_i\xi_i.
\label{eq:pca-weights}
\end{equation}
Weighted principal component analysis estimates $d$ local directions:
\begin{equation}
T\in\operatorname*{arg\,min}_{T^\top T=I_d}
\sum_i\pi_i\|(I-TT^\top)(\xi_i-c)\|^2,
\qquad P_T=TT^\top,\quad P_N=I-P_T.
\label{eq:fit}
\end{equation}
{For each head-specific refit, we replace $\pi_i$ by that head's normalized weights and repeat the decomposition after recentering.}

\subsubsection{CMAT cylinder scales and support}
{For CMAT, we use a \emph{single finite cylinder} to restrict attention to a subset of the retrieved observations, with baseline scales}
\begin{equation}\begin{aligned}
\widehat\sigma&=\operatorname{clip}\left\{\frac{\sqrt{\sum_i\pi_i\|P_N(\xi_i-c)\|^2/\max(p-d,1)}}{\ell},10^{-5},0.95\right\},\\
R_0&=\frac{5\widehat\sigma\ell}{\log_{10}K},\qquad
H_0=\frac{10\widehat\sigma\sqrt{\log(1/\widehat\sigma)}\ell}{\log_{10}K}.
\end{aligned}\label{eq:kernel-scales}\end{equation}
{Here $R_0$ and $H_0$ denote the baseline cylinder radius and axial half-length, and $K$ is the number of retrieved comparison objects.}

Let $d_{(j)}$ be the $j$th ordered value of $\|\xi_i\|$. A preliminary spherical average defines the cylinder direction:
\begin{equation}
\begin{aligned}
r_{\rm sph}&=\max\{R_0,1.10d_{(\min(5,K))},10^{-12}\},\\
\omega_i&=[1-(\|\xi_i\|/r_{\rm sph})^2]_+^3,\qquad
m_0=\frac{\sum_i\omega_i\xi_i}{\sum_i\omega_i},\qquad a=m_0/\|m_0\|.
\end{aligned}\label{eq:cylinder-axis}
\end{equation}
{If $\|m_0\|\leq10^{-12}\ell$, we use the first fitted direction as the support axis; otherwise the axis is given by $a$ above. Its axial and radial coordinates are $u_i=a^\top\xi_i$ and $v_i=\|\xi_i-u_i a\|$, even when $a$ is neither a tangent nor a normal direction. We require at least $m=\min(24,K)$ comparison objects for correction and $m=\min\{\max(11,d+2),K\}$ for generation, expanding the support as follows:}
\begin{equation}\begin{aligned}
z_i&=\max\{v_i/R_0,|u_i|/H_0\},\qquad e=\max\{1,1.10z_{(m)}\},\\
R&=eR_0,\qquad H=eH_0,\qquad
\mathcal C=\{i:v_i<R,\ |u_i|<H\}.
\end{aligned}\label{eq:cylinder-support}\end{equation}
{The radius and half-length expand by the same factor $e$, giving a total cylinder length of $2H$ and a common support for all four heads. For generation with $d=12$, the minimum support is $m=\min(14,K)$; for the YY-based comparator, we use the uniform-weight construction in Appendix~\ref{app:yy-kernels}.}

\subsection{Weights and update size for generation}\label{app:cmat-core-details}

{For the four fixed geometric heads used in generation, we express offsets relative to the preliminary basis $T$ as $t_i=T^\top\xi_i$ and $n_i=P_N\xi_i$. The distance scales $s_T^2,s_N^2,s_D^2$ are the medians of $\|t_i\|^2,\|n_i\|^2,\|\xi_i\|^2$, respectively, each floored at $10^{-8}\ell^2$, giving the scores and weights}
\begin{equation}\begin{aligned}
L_{ih}&=-a_h\frac{\|t_i\|^2}{s_T^2}-b_h\frac{\|n_i\|^2}{s_N^2}-q_h\frac{\|\xi_i\|^2}{s_D^2},\\
\alpha_{ih}&=\frac{\mathbf1\{i\in\mathcal C\}\exp L_{ih}}
{\sum_{j\in\mathcal C}\exp L_{jh}}.
\end{aligned}\label{eq:kernel}\end{equation}
{We compute all median distance scales and the preliminary tangent basis from the full retrieved neighborhood, sharing this basis and the cylinder across the four heads with coefficients given in Table~\ref{tab:heads}. For numerical stability, we subtract the maximum supported score before exponentiation and set all weights outside the cylinder to zero.}

{We average the head-specific increments $h_h=\sum_i\alpha_{ih}\xi_i$ to obtain $h=\tfrac14\sum_{h'=1}^4h_{h'}$, with mean attention weights $\bar\alpha_i=\tfrac14\sum_{h'=1}^4\alpha_{ih'}$. The step parameters in~\textup{(2.27)} of the main article are}
\begin{equation}
g=0.70\min\left\{1,\frac{1}{(d+2)\sum_i\bar\alpha_i^2}\right\},\qquad \eta=0.10.
\end{equation}
{For $d=12$, an effective support of $1/\sum_i\bar\alpha_i^2\geq14$ gives the maximum step factor $g=0.70$, provided the local singular-value check also passes. If the $d$th singular value is at most $10^{-10}\max(\sigma_1,10^{-12})$, we instead set $g=0$ regardless of effective support.}

\Needspace{6\baselineskip}

\subsection{Computational details of rare-pose generation}\label{app:generation-details}
\subsubsection{Selecting rare starting poses and estimating neighborhood scale}
{We first screen 250,000 search poses using subject-excluded calibration and $\rho_{g,-h}\geq0.95$, retaining subjects with at least two rare anchors. In each attempt, we sample an anchor uniformly with replacement and then a partner uniformly from its at most 12 nearest other rare anchors of the same subject under product rotation distance.}

{To fit the local geometry using the complete reference, we retrieve the 65 nearest reference poses, discard the closest returned entry, and form $\mathcal L_a$ from the remaining 64. Since the anchor belongs to the reference, a zero-distance match is available, although tied entries with identical rotations mean that the discarded entry need not be the anchor's own record. We then form the weighted centered matrix of relative rotations, retain its first 12 right singular vectors, and set $\ell_a=\sqrt{\max(\operatorname{median}_i\|\xi_i\|^2,10^{-16})}$.}

\subsubsection{Generating candidates along observed variation and bounding changes}
{For the same anchor and partner, we generate six candidates with independent draws of $\tau_b,z_b$ in main-paper equation~\textup{(2.25)}, using the local singular values to scale exploration along the fitted directions.}

{For $v\in\mathbb R^{63}$, we bound the total length before capping the individual joint rotations:}
\begin{equation}\begin{aligned}
 v'&=v\min\{1,0.65\ell_a/\max(\|v\|,10^{-12})\},\\
 [\mathcal B_a(v)]_j&=v'_j\min\{1,(12\pi/180)/\max(\|v'_j\|,10^{-15})\}.
\end{aligned}\label{eq:global-cap}\end{equation}
{After fitting the geometry at the initial candidate, we compute its CMAT increment, cap it at $12^\circ$ per joint, and compose it with the candidate rotations. We then recompute the total principal relative rotation vector from the anchor and apply $\mathcal B_a$, so that the anchor-relative bound limits the actual angular displacement after composition.}

\subsubsection{Comparing rarity, reference distance and local deviation}

{For an anchor from subject $h$ and collection $g$, let $\mathcal C_g,w_i$ denote the independent calibration set and weights defined in Section~2.2 of the main article. Holding $\mathcal L_a$ fixed, we refit the local geometry at each $q_b$ and calculate}
\begin{equation}\begin{aligned}
\widehat Q^{(g)}_\alpha(z)&=\inf\left\{u:\sum_{i\in\mathcal C_g}w_i\mathbf1\{z_i\leq u\}\geq\alpha\right\},\\
\rho_b&=\rho_{g,-h}(q_b),\qquad D_b=\min_{r\in\mathcal R}d(q_b,r),\\
\nu_g&=\max\left\{\widehat Q^{(g)}_{0.01}
\left(\left\{\min_{r\in\mathcal R}d(c_i,r)\right\}_{i\in\mathcal C_g}\right),10^{-6}\right\},\\
E_b&=E(q_b;\mathcal L_a),\qquad
q_b\in\mathcal R\ \Longrightarrow\ D_b=0<\nu_g.
\end{aligned}\label{eq:generation-candidate-quantities}\end{equation}
\begin{equation}\begin{aligned}
\mathcal B_{\rm ok}&=\{b\in\{1,\ldots,6\}:\rho_b\geq0.95,\ D_b>\nu_g,\ E_b\leq0.55\},\\
b_*&=\begin{cases}
\min\bigl(\operatorname*{arg\,min}_{b\in\mathcal B_{\rm ok}}E_b\bigr),&\mathcal B_{\rm ok}\ne\varnothing,\\
1,&\mathcal B_{\rm ok}=\varnothing,
\end{cases}\qquad q_c=q_{b_*}.
\end{aligned}\label{eq:generation-candidate-selection}\end{equation}
{We refine $q_c$ before the final acceptance checks in Appendix~\ref{app:generation-screening}.}

\subsubsection{Refining a candidate while preserving rarity}\label{app:generation-refinement}

{For refinement, we retrieve 64 neighbors of the selected candidate $q_c$ from the complete reference to form $\mathcal L_c$, compute one CMAT increment, and cap it at $12^\circ$ per joint to obtain $\delta_c$. We consider the trials}
\begin{equation}
q(u)=\operatorname{Exp}_{q_c}(u\delta_c),\qquad u\in\{0,0.25,0.5,0.75,1\}.
\label{eq:search}
\end{equation}
{At each trial, we calculate local deviation using the fixed $\mathcal L_c$ while updating rarity and reference distance, requiring both the three candidate conditions and}
\begin{equation}
\rho_{g,-h}\{q(u)\}\geq\rho_{g,-h}(q_c)-0.005.
\label{eq:rollback}
\end{equation}
{Among feasible trials, we choose the one with the smallest $E$, breaking ties by the displayed order. If none is feasible, we retain the zero step for final screening.}

\subsubsection{Final checks and yield}\label{app:generation-screening}
{At the final pose, we retrieve 64 neighbors anew from the complete reference and recompute rarity, separation and local deviation before applying~\textup{(2.24)} of the main article together with the following joint-based checks.}

\begin{equation}\begin{aligned}
P(q)&=[p_0(q),\ldots,p_{21}(q)]^\top
=\operatorname{SMPL}_{0:21}(q;\beta_{\rm shape}=0),\qquad p_{\rm root}=p_0,\\
\operatorname{isfinite}(q_f)&\equiv1,\qquad
\operatorname{isfinite}\{P(q_f)\}\equiv1.
\end{aligned}\label{eq:generation-finite}\end{equation}
\begin{equation}\begin{split}
\mathcal P=\{&(4,7),(5,8),(7,10),(8,11),(10,11),(15,16),(15,20),(16,19),\\
&(17,18),(19,20),(4,15),(5,16),(7,17),(8,18),(10,19),(11,20)\}.
\end{split}\label{eq:generation-joint-pairs}\end{equation}
\begin{equation}\begin{aligned}
s(q)&=\max\{\|p_2(q)-p_1(q)\|+\|p_3(q)-p_2(q)\|,10^{-6}\},\\
d^{J}_{uv}(q)&=\frac{\|p_u(q)-p_v(q)\|}{s(q)},\\
l_{g,uv}&=\max\left\{\widehat Q^{(g)}_{0.002}
\bigl(\{d^{J}_{uv}(c_i)\}_{i\in\mathcal C_g}\bigr)-0.002,0\right\},\\
d^{J}_{uv}(q_f)&\geq l_{g,uv}\qquad\text{for all }(u,v)\in\mathcal P.
\end{aligned}\label{eq:generation-joint-bounds}\end{equation}

{Only poses that pass all the checks above are accepted, and we calculate yield as defined in Appendix~\ref{app:generation-yield}.}

\subsubsection{Retrieval settings and verification results}\label{app:generation-retrieval}
{For rarity retrieval, we project rotation-matrix embeddings to 64 PCA coordinates, screen candidates with an IVF index, and rerank them by product rotation distance. The index has 1,024 cells, of which each query probes 64 to retrieve 2,048 frames. If fewer than 16 distinct sequences remain after subject exclusion, we expand the candidate count successively to 8,192, 32,768 and 65,536. We use this configuration for both calibration and queries, explicitly converting between scalar-first quaternions for geometry and scalar-last quaternions for rarity retrieval.}

Checks on fixed poses gave the following results:
\begin{enumerate}
\item Rebuilding the index left rarity labels and percentiles unchanged for 3,302 evaluation poses.
\item Exhaustive queries for 16 boundary or accepted poses gave a maximum score difference of $5.40\times10^{-4}$. One rarity label changed; separation labels did not. This check retained the saved calibration scores.
\item Increasing cells probed / candidate frames from 64/2,048 to 128/8,192 and recalibrating changed one rarity label among 12,288 fixed outputs (2,048 per generator). All acceptance decisions were unchanged; CMAT still accepted 705.
\end{enumerate}
These checks did not rerun candidate selection; the expanded search remains approximate. Methods in the main comparison share screening conditions, but their internal search costs are not matched.

\section{Controlled comparisons and their scope}\label{app:controlled-supp}

\subsection{Definition and scope of the motion-bound analysis}\label{app:bounded-noise}

Let $P_{itj}$ be the clean root-relative coordinate of joint $j$ in frame $t$ of clip $i$, and let $Y_{irtj}$ be its noisy counterpart in repetition $r\in\{1,2\}$. Each clip contains $T=240$ frames at 30 Hz and $J=22$ joints. Define the clean-target body scale by
\begin{equation}
s_i=\operatorname{median}_{t}
\left(\frac1J\sum_{j=1}^J\|P_{itj}\|^2\right)^{1/2}.
\end{equation}
The motion descriptors are computed separately for each noisy repetition:
\begin{equation}\begin{aligned}
V_{ir}&=\frac{30}{s_iJ(T-1)}\sum_{t=2}^T\sum_{j=1}^J
\|Y_{irtj}-Y_{ir,t-1,j}\|,\\
A_{ir}&=\frac1{s_i}\left\{\frac1{JT}\sum_{t,j}
\|Y_{irtj}-\bar Y_{irj}\|^2\right\}^{1/2},\qquad
\bar Y_{irj}=\frac1T\sum_tY_{irtj}.
\end{aligned}\label{eq:bounded-descriptors}\end{equation}
{Here $V_{ir}$ is speed in body scales per second and $A_{ir}$ is temporal amplitude in body scales.}

The cached descriptors are $\bar V_i=(V_{i1}+V_{i2})/2$ and $\bar A_i=(A_{i1}+A_{i2})/2$. With equal weight for each of the 236 clips and both sources pooled, let $Q_q$ be an empirical quantile calculated by linear interpolation. The selected rule is
\begin{equation}\begin{aligned}
c_V&=\frac{\lceil100Q_{1.00}(\bar V)\rceil}{100}=6.34,\\
c_A&=\frac{\lceil100Q_{0.90}(\bar A)\rceil}{100}=0.73,\\
I_i&=\mathbf1\{\bar V_i\leq c_V,\ \bar A_i\leq c_A\}.
\end{aligned}\label{eq:bounded-rule}\end{equation}

\subsection{All component comparisons on the retained population}\label{app:bounded-ablation}

\subsubsection{Definitions of component interventions}\label{app:correction-ablation}
{Each comparison changes one operation or an explicitly specified group of operations in the frozen correction algorithm. Except for variants that change candidate selection, all use $f=0.8$, $\eta_c=30$, the best-candidate policy, the same two noise realizations and the same numerical budget. Evaluation uses the clips retained in Appendix~\ref{app:bounded-noise} and the same rare-frame labels.}

\begin{enumerate}
\item \emph{No preliminary guide}: use the normalized raw observations for matching and fitting, set neighbor reliability and the smoothing-retention factor to one, and estimate working variance for the unsmoothed input. Retain the final temporal term.
\item \emph{No cylinder boundary}: let all retrieved neighbors participate in the weight calculations; retain their distance scores and other operations.
\item {\emph{Reliability-only weighting}: replace all four rules by weights proportional only to neighbor reliability within the cylinder. This jointly removes distance decay, balance and GLS weighting; it does not isolate distance weighting.}
\item \emph{No reliability weighting}: set neighbor reliability to one while retaining the cylinder and the remaining weight calculations.
\item \emph{No balance head}: remove its four projection candidates and reselect among the remainder.
\item \emph{No GLS head}: remove its four projection candidates and reselect among the remainder.
\item \emph{No balance or GLS heads}: retain only the two distance-based weight rules and their projections, and reselect.
{\item \emph{Three free tangent directions only}: all four weighting rules are retained, but only the constraint projection $I-\sum_{k=1}^{3}b_kb_k^\top$ is permitted, leaving three tangent directions free. This is the variant labeled ``Normal projection only'' in the component tables.}
\item \emph{Full-center constraint only}: retain all four weight rules, but permit only the identity projection $I$.
\item \emph{No displacement coefficient}: set $c_{gh}=1$ in both proposals and final geometric terms; recompute the candidate sensitivity scores.
\item \emph{No null option}: choose the best candidate even when all penalized scores are nonnegative.
\item \emph{No score margin}: omit the standard-error term from the candidate score, retaining the mean score and null option.
\item \emph{Equal candidate mixture}: assign $1/16$ to every candidate, without score-based selection or rejection.
\item \emph{No final temporal penalty}: set $\lambda_t=0$ only in the final objective; retain the full method's guide and geometric constraints.
\item \emph{Quadratic final observation loss}: replace only the final robust observation term by squared error.
\item \emph{No overlap normalization}: set $\gamma_g=1$ in the geometric objective.
\item \emph{Temporal neighbors}: replace feature matching by the nearest 32 nonself windows in time.
\item \emph{Position-only matching}: omit first and second differences from the retrieval features, retaining the same neighbor-count and temporal-separation rules.
\item \emph{Temporal-only control}: remove all local geometric terms and retain the common robust observation and temporal objective.
\end{enumerate}

\begin{table}[!htbp]\centering\small
\setlength{\tabcolsep}{5pt}
\caption{Complete component comparison on the selected exploratory population: MPJPE (mm). Entries are variant minus full CMAT; positive changes favor full CMAT. }\label{tab:bounded-full-mpjpe}
\begin{tabular}{lrrr}\hline
Variant & Change & 95\% interval & $q$\\\hline
Temporal terms only & +11.663 & $[+10.183,+13.102]$ & $<0.001$\\
No preliminary guide & -0.774 & $[-1.195,-0.373]$ & $<0.001$\\
No cylinder boundary & -0.235 & $[-0.614,+0.153]$ & 0.271\\
{Reliability-only weights} & +1.112 & $[+0.852,+1.381]$ & $<0.001$\\
No reliability factor & -0.036 & $[-0.103,+0.038]$ & 0.351\\
No balance head & +0.168 & $[+0.005,+0.364]$ & 0.091\\
No GLS head & +0.838 & $[+0.530,+1.154]$ & $<0.001$\\
No balance or GLS heads & +0.974 & $[+0.644,+1.311]$ & $<0.001$\\
Normal projection only & +7.267 & $[+6.220,+8.268]$ & $<0.001$\\
Full-center constraint only & +0.161 & $[-0.075,+0.416]$ & 0.240\\
No displacement coefficient & +0.153 & $[+0.053,+0.265]$ & 0.007\\
No null candidate & +0.005 & $[0.000,+0.014]$ & 0.225\\
No score standard-error term & -0.042 & $[-0.218,+0.139]$ & 0.672\\
Equal candidate mixture & +1.992 & $[+1.203,+2.775]$ & $<0.001$\\
No final temporal penalty & +0.157 & $[+0.141,+0.175]$ & $<0.001$\\
Quadratic final observation loss & -0.124 & $[-0.172,-0.082]$ & $<0.001$\\
No overlap normalization & +0.258 & $[+0.130,+0.394]$ & $<0.001$\\
Nearest temporal neighbors & +2.409 & $[+1.699,+3.157]$ & $<0.001$\\
Position-only matching & +0.485 & $[+0.134,+0.831]$ & 0.009\\
\hline\end{tabular}
\end{table}

\begin{table}[!htbp]\centering\small
\setlength{\tabcolsep}{5pt}
\caption{Complete component comparison on the selected exploratory population: PA-MPJPE (mm). Entries are variant minus full CMAT; positive changes favor full CMAT. }\label{tab:bounded-full-pampjpe}
\begin{tabular}{lrrr}\hline
Variant & Change & 95\% interval & $q$\\\hline
Temporal terms only & +9.510 & $[+7.858,+10.998]$ & $<0.001$\\
No preliminary guide & -0.688 & $[-1.013,-0.365]$ & $<0.001$\\
No cylinder boundary & -0.254 & $[-0.618,+0.107]$ & 0.220\\
{Reliability-only weights} & +0.872 & $[+0.634,+1.104]$ & $<0.001$\\
No reliability factor & -0.049 & $[-0.089,-0.011]$ & 0.019\\
No balance head & +0.126 & $[-0.004,+0.275]$ & 0.101\\
No GLS head & +0.582 & $[+0.298,+0.867]$ & $<0.001$\\
No balance or GLS heads & +0.692 & $[+0.400,+0.985]$ & $<0.001$\\
Normal projection only & +6.047 & $[+4.981,+7.037]$ & $<0.001$\\
Full-center constraint only & +0.003 & $[-0.173,+0.193]$ & 0.972\\
No displacement coefficient & +0.122 & $[+0.041,+0.206]$ & 0.006\\
No null candidate & +0.004 & $[0.000,+0.012]$ & 0.224\\
No score standard-error term & -0.063 & $[-0.219,+0.091]$ & 0.448\\
Equal candidate mixture & +1.655 & $[+0.866,+2.402]$ & $<0.001$\\
No final temporal penalty & +0.118 & $[+0.099,+0.135]$ & $<0.001$\\
Quadratic final observation loss & -0.092 & $[-0.144,-0.049]$ & $<0.001$\\
No overlap normalization & +0.207 & $[+0.080,+0.349]$ & 0.004\\
Nearest temporal neighbors & +1.921 & $[+1.538,+2.280]$ & $<0.001$\\
Position-only matching & +0.489 & $[+0.224,+0.744]$ & $<0.001$\\
\hline\end{tabular}
\end{table}

\begin{table}[!htbp]\centering\small
\setlength{\tabcolsep}{5pt}
\caption{Complete component comparison on the selected exploratory population: acceleration error (mm/frame$^2$). Entries are variant minus full CMAT; positive changes favor full CMAT. }\label{tab:bounded-full-acceleration}
\begin{tabular}{lrrr}\hline
Variant & Change & 95\% interval & $q$\\\hline
Temporal terms only & +1.139 & $[+0.958,+1.294]$ & $<0.001$\\
No preliminary guide & +0.061 & $[+0.028,+0.091]$ & $<0.001$\\
No cylinder boundary & -0.007 & $[-0.023,+0.009]$ & 0.431\\
{Reliability-only weights} & +0.004 & $[-0.016,+0.022]$ & 0.723\\
No reliability factor & +0.002 & $[-0.001,+0.006]$ & 0.306\\
No balance head & -0.008 & $[-0.026,+0.007]$ & 0.358\\
No GLS head & -0.035 & $[-0.044,-0.026]$ & $<0.001$\\
No balance or GLS heads & -0.046 & $[-0.066,-0.028]$ & $<0.001$\\
Normal projection only & +0.246 & $[+0.194,+0.298]$ & $<0.001$\\
Full-center constraint only & +0.020 & $[+0.004,+0.038]$ & 0.029\\
No displacement coefficient & +0.055 & $[+0.046,+0.066]$ & $<0.001$\\
No null candidate & +0.002 & $[0.000,+0.006]$ & 0.296\\
No score standard-error term & +0.009 & $[0.000,+0.018]$ & 0.059\\
Equal candidate mixture & -0.055 & $[-0.094,-0.021]$ & 0.005\\
No final temporal penalty & +6.900 & $[+6.642,+7.198]$ & $<0.001$\\
Quadratic final observation loss & -0.008 & $[-0.014,-0.002]$ & 0.015\\
No overlap normalization & +0.224 & $[+0.203,+0.247]$ & $<0.001$\\
Nearest temporal neighbors & -0.032 & $[-0.052,-0.010]$ & 0.004\\
Position-only matching & -0.029 & $[-0.046,-0.012]$ & 0.002\\
\hline\end{tabular}
\end{table}

\FloatBarrier

\section{Correction of 3DPW--EFT pose estimates}\label{app:aist-correction}\label{sec:real-correction}
{To evaluate correction of real video-based pose estimates, we use 3DPW--EFT data comprising 35,515 frames in 37 person tracks from 24 videos. Because rarity screening leaves few frames for a separate tail analysis, we evaluate correction over all frames, computing spatial errors for 14 joints relative to the hip-midpoint root and acceleration errors at the 34,752 valid centers with consecutive original frame identifiers.}

\begin{table}[!htbp]\centering\small
\caption{All-frame 3DPW--EFT errors: 35,515 frames and 34,752 valid acceleration centers from 37 person tracks.}\label{tab:eft-all}
\begin{tabular}{lrrr}\hline
Method & MPJPE (mm) & PA-MPJPE (mm) & Acc. (mm/frame$^2$)\\\hline
Original estimate & 90.324 & 52.194 & 31.073\\
YY-based & 87.829 & 50.620 & 5.087\\
SmoothNet & 88.403 & 51.989 & 5.105\\
Self-Action CMAT & 87.819 & 50.611 & 5.088\\
\hline\end{tabular}
\end{table}

\section{The kernel functions in Yao--Yau fitting}\label{app:additional-geometry}\label{app:yy-kernels}

The full kernel functions used in~\textup{(2.11)} of the main article are
\begin{equation}\begin{aligned}
w_{\parallel}(u)&=\begin{cases}
1,&\|u\|\leq H/2,\\
[1-\{(2\|u\|-H)/H\}^2]^k,&H/2<\|u\|<H,\\
0,&\|u\|\geq H,
\end{cases}\\
w_{\perp}(v)&=[1-\|v\|^2/R^2]_+^k,\qquad k>2.
\end{aligned}\label{eq:yao-kernels}\end{equation}
{Whereas~\eqref{eq:yao-kernels} specifies the original smooth kernels, our implemented YY-based comparator uses uniform means over the cylindrical support constructed below.}

{For a requested support size $m$, let $d_{(j)}$ denote the ordered query distances and \[j=\min\{K,\max(5,\lceil m/2\rceil)\}.\] Using $R_0,H_0$ from~\eqref{eq:kernel-scales}, we set the pilot sphere radius to the larger of $2R_0$ and the smallest representable value above $d_{(j)}$. We define the cylinder axis from the unweighted mean within that sphere, using the first fitted direction when the mean is zero.}

{We compute the ratios in~\eqref{eq:cylinder-support} and enlarge both scales by the larger of one and the smallest representable value above $z_{(m)}$, retaining boundary ties. Finally, we assign each supported observation weight $1/|\mathcal C|$ and refit the centers and directions.}

\FloatBarrier
\bibliographystyle{cmat-plainnat}
\bibliography{references}